# Pits, Uplifts and Small Chaos Features on Europa: Morphologic and Morphometric Evidence for Intrusive Upwelling and Lower Limits to Ice Shell Thickness

Kelsi N. Singer[1], William B. McKinnon[2], and Paul M. Schenk[3].

[1]Southwest Research Institute, 1050 Walnut St. Suite 300, Boulder, CO 80302
ksinger@levee.wustl.edu
[2]Department of Earth and Planetary Sciences and McDonnell Center for the Space Sciences, Washington University in St. Louis, 1 Brookings Dr., Saint Louis, MO 63130, USA
mckinnon@wustl.edu
[3]Lunar and Planetary Institute, Houston, TX 77058.
schenk@lpi.usra.edu

## Abstract

One of the clearest but unresolved questions for Europa is the thickness of its icy shell. Europa's surface is resplendent with geological features that bear on this question, and ultimately on its interior, geological history, and astrobiological potential. We characterize the size and topographic expression of circular and subcircular features created by endogenic thermal and tectonic disturbances on Europa: pits, uplifts, and small, subcircular chaos. We utilize the medium-resolution Galileo regional maps (RegMaps), as well as high-resolution regions, digital elevation models derived from albedo-controlled photoclinometry, and in some cases stereo-controlled photoclinometry. While limited in extent, the high-resolution images are extremely valuable for detecting smaller features and for overall geomorphological analysis. A peak in the size-distribution for all features is found at ~5–6 km in diameter and no pits smaller than 3.3 km in diameter were found in high resolution images. Additionally, there is a trend for larger pits to be deeper, and larger uplifts to be higher. Our data support a diapiric or intracrustal sill interpretation (as opposed to purely non-intrusive, melt-through models) and place a *lower limit* on ice shell thickness at the time of feature formation of 3-to-8 km, assuming isostasy and depending on the composition of the ice and underlying ocean.

# 1 Introduction

Constraining the thickness of the ice shell on Europa and the geological processes occurring in it are a key to understanding this icy world and its potential habitability (e.g., Collins and Nimmo, 2009; Greeley et al., 2009; Hand et al., 2009; McKinnon et al., 2009; Nimmo and Manga, 2009; Peddinti and McNamara, 2019; Howell et al., 2020). We focus on features generally agreed to have been created by endogenic processes in Europa's ice shell or ocean: pits, uplifts, and subcircular chaos. Pits and uplifts are defined by their negative or positive topographic expression, respectively. Pits and uplifts generally retain pre-existing surface structures such as ridges, while chaos specifically refers to areas where the surface is broken up, in some cases to the point of destroying all original surface topography. Although circular, impacts are an exogenic geologic process and thus not considered here.

Inspired by the data returned by the *Galileo* spacecraft mission, several research groups have mapped these features and came to different conclusions about their formation mechanism. Greenberg et al. (2003) mapped sub-circular features with clear topographic expressions, but explicitly left out chaos regions previously mapped (Greenberg et al., 1999; Riley et al., 2000). Greenberg et al. (2003) concluded that a continuous range of feature sizes existed, with exponentially increasing numbers of small features, down to the resolution limit. This distribution was considered to be in support of a very thin ice shell (less than a few km thick) and a "melt-through" model of feature formation, where localized heating thins the shell, exposing water below. Near-equatorial mapping by Spaun et al. (2001; 2004) and Spaun (2002) focused on large and small chaos regions, but also noted pits and other features. Their mapping found a peak in size distribution near ~5 km and declining numbers of smaller features. This distribution was interpreted as an indication of a diapiric formation model, which implies a thicker ice shell heated from below that induces the buoyant rise of warmer ice masses (Pappalardo et al., 1998; Pappalardo and Barr, 2004; Barr and

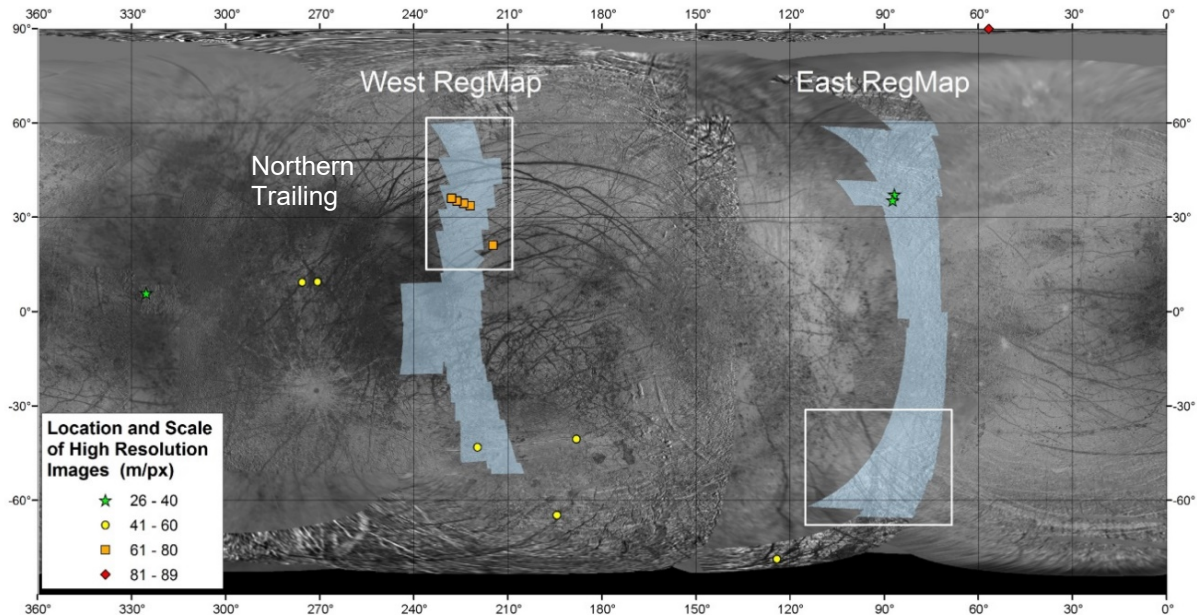

**Figure 1: Simple cylindrical projection of Europa with summary of mapped locations.** Shading indicates the East (leading) and West (trailing) regional maps (RegMaps, ~220 and 210 m px$^{-1}$, respectively) and symbols show locations of higher resolution images with features. Boxes outline areas richest in mapped features (pits, uplifts and sub-circular small chaos regions). The remaining areas have a high percentage of continuous, or essentially interconnected, chaos. Basemap courtesy USGS.

Showman, 2009). These two hypotheses (melt through or diapiric feature formation) were expanded upon and many others have been explored in later years (e.g., brine mobilization, sill intrusion, cryovolcanic processes, the role of impacts, and hybrid models). The relative merits of different formation mechanism hypotheses for chaotic terrain and other upwelling features have been discussed in many sources (e.g., Pappalardo et al., 1998; Collins et al., 2000; Fagents, 2003; Goodman et al., 2004; Nimmo et al., 2004; Pappalardo and Barr, 2004; Collins and Nimmo, 2009; Schmidt et al., 2011; Kattenhorn and Prockter, 2014; Michaut and Manga, 2014; Cox and Bauer, 2015; Craft et al., 2016; Culha and Manga, 2016; Manga and Michaut, 2017; Quick et al., 2017; Craft et al., 2019; Noviello et al., 2019; Buffo et al., 2020; Lesage et al., 2020; Steinbrügge et al., 2020).

The goal of this work is to introduce more detailed information about feature topography to the discussion, building on several earlier studies (Schenk and McKinnon, 2001; Schenk and Pappalardo, 2004; Singer et al., 2010; Singer, 2013). We have mapped all circular and subcircular features plausibly created by upwellings or other endogenic processes in the size range of 1 to 50 km in diameter and incorporated previously unavailable topographic data as an aid to mapping and characterization of features. Results of this new mapping show decreasing numbers of small features, and a peak in the size distribution for all features at ~5–6 km in diameter. Topography was also used to find the depths and heights of pits and uplifts in the mapped regions. A general trend of increasing pit depth with increasing pit size was found, a correlation more easily understood in the context of a diapiric or sill intrusion hypothesis for feature formation.

The paper proceeds as follows. Mapping techniques are summarized in section 2, and the results are given in section 3. Section 4 compares the results to those of previous studies and discusses the implications of these results for formation mechanisms and ice shell thickness. A summary of our main findings and outlook for future work are given in section 5.

# 2 Subcircular Endogenic Features on Europa

## 2.1 Mapping Methods

Our mapping was conducted over the two large, N-S mosaics taken during the Galileo mission's regional mapping campaign (East and West RegMaps) and over all images at higher resolutions (~6-to-200 m px$^{-1}$). Figure 1 provides an overview of mapped locations, and Fig. 2 displays features mapped on the RegMaps. Base image mosaic resolutions are 220 and 210 m px$^{-1}$ for the East and West RegMaps, respectively. The vertical (north-south) resolution of the original RegMap images decreases somewhat poleward, as Galileo viewed the surface more obliquely at higher latitudes (up to ~440 m px$^{-1}$ at the greatest poleward extents). The total area of the RegMaps mapped here covers ~9% of the surface of Europa. Two areas are the richest in circular and subcircular features: the southern section of the East RegMap (leading hemisphere; Fig 2b) and the northern section of the West RegMap (trailing hemisphere; Fig 2c).

Features were identified based on morphology and further classified based on topography (Figs. 3-4). All circular to subcircular features were mapped, including pits, uplifts and chaos. Impact craters were mapped as part of this process but not measured or reported on here. We did not map all incidents of non-circular chaos, as previous studies have focused on these (Riley et al., 2000; Spaun, 2002; Figueredo and Greeley, 2004), but instead demarcated large expanses of essentially interconnected chaos as "regional chaos." See work on global geologic mapping for information on larger extents of chaotic terrain (e.g., Greeley et al., 2000; Doggett et al., 2009; Leonard et al., 2017; Senske et al., 2018). Individual feature types are defined in this paper as follows:
- *Pits* are circular to elliptical features with a negative topographic expression and little to no apparent surface disruption (pre-existing surface ridge structures are largely preserved).

- *Uplifts* are subcircular features with a positive topographic expression and little to no apparent surface disruption.
- *Subcircular chaos regions* are surface areas with a hummocky disrupted texture distinct from their ridged surroundings (including a few high standing tilted blocks that appear to be lone peaks at the current resolution but may be part of chaos regions). Some, but not all, are associated with a darker albedo compared to the surrounding plains.
- *Spots* are generally defined as low albedo areas with no discernable topographic expression and little to no apparent surface disruption. Any features that would have had this classification in the RegMaps were small, and deemed likely to be chaos with texture too small to be discerned in the resolution limit. Only a few features in the RegMaps fell into this category, and were classified with chaos for this study.

All of these features likely lie in a continuum of surface expressions from variations on a similar mechanism of subsurface upwelling (see summary in Collins and Nimmo, 2009), rather than each feature type being created by a substantially different subsurface processes. For the purposes of utilizing the topographic data, the features are defined here as separate units. A related, but different, concept is that features may have a typical progression from one type of feature to another, e.g., features that start out as pits eventually become uplifts or chaos later in their evolution (Sotin et al., 2002; Schmidt et al., 2011; Manga and Michaut, 2017). At the available image resolutions, there are no strong morphological indicators showing that features progress from one type to another, but smaller-scale fractures or other signs (perhaps along the feature boundaries) may be searched for with future, higher-resolution data.

The dark, subcircular chaos features in this study would have been called lenticulae (a broad, nongenetic term stemming from the Latin word for freckles) when viewed in lower resolution Voyager images. Because the word "lenticulae" implies an albedo lower than the surroundings, and many of the features mapped in this study do not have this characteristic, we did not use the term here to describe the entire feature set. The basic morphological categories defined here are similar to Greenberg et al. (2003; see their figure 7), but here we also use the topography and additional criteria to map features, as described below.

We identified features visually, and then used topographic profiles to confirm and define their extent. For pits and uplifts, the inversion point in the topography—where the slope transitions from the steep slope of the feature wall/side, into the shallow-to-flat slope for the surrounding terrain—was selected as the feature edge where possible. The edges of each feature were tied to inversion points in the topography with four to six profiles (examples in Fig. 4). Partial features (those which appear to continue outside of the image boundary) are not included in the results below. Depressions bounded on all sides by ridges (as in, the pit-like topography could be due simply to the fact that ridges are raised features) were not included. Appendix A discusses in detail what features were included, or not, in the data set. Appendix A also describes how the number of features changes when using a more (or less) conservative feature definition, but the basic results given in the next section remain the same.

## 2.2 High Resolution Mapping

All Galileo images with resolutions between ~6 and 200 m px$^{-1}$ were searched for the features described above. Images where features were found ranged from 26-to-100 m px$^{-1}$ and had varied lighting geometries (phase angles from ~30° to 119°). Although limited in extent, high resolution images complement the moderate resolution RegMaps. High resolution images allow a search for features smaller than would be discernable on the RegMaps (discussed further below) and permit a more detailed examination of feature morphology. Example features in Fig. 5 illustrate how some pits clearly preserve preexisting ridged terrain. Even the very limited, high resolution image data set available for Europa exhibits many different variations on feature morphologies. At high resolution the line between uplifts and chaos is blurrier, as many uplifts are crossed by at least one fracture (as might be expected from the tensile stresses resulting from extension of the surface during uplift). From our study of high-resolution images, only one smooth (at the image resolution), sub-circular feature was noted (Fig. 5d). This feature has been identified in previous studies (e.g., Head et al., 1998; Fagents, 2003), and has been interpreted as an embayment of fluid-like material from below based on its morphology and contacts with surrounding terrains. Although there are other relatively smooth patches, no other sub-circular candidates similar to Fig. 5d were found in the available images.

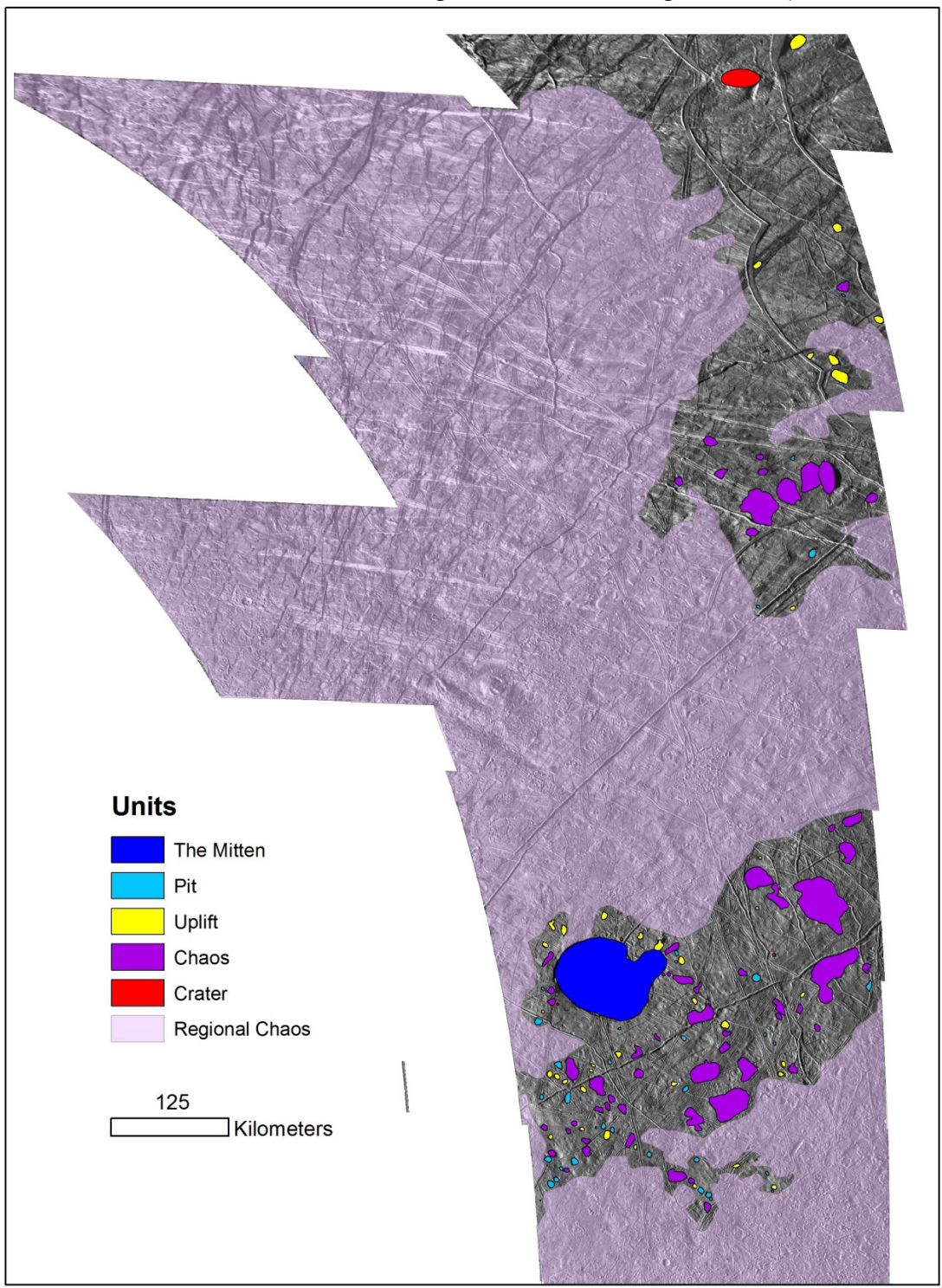

**Figure 2a: East RegMap – Northern leading region.** Scale bar applies to the latitude at which it is placed. Extent: 10° to 61°N and 70° to 107°W. Mapped area (not including the regional chaos) = $1.8 \times 10^5$ km$^2$.

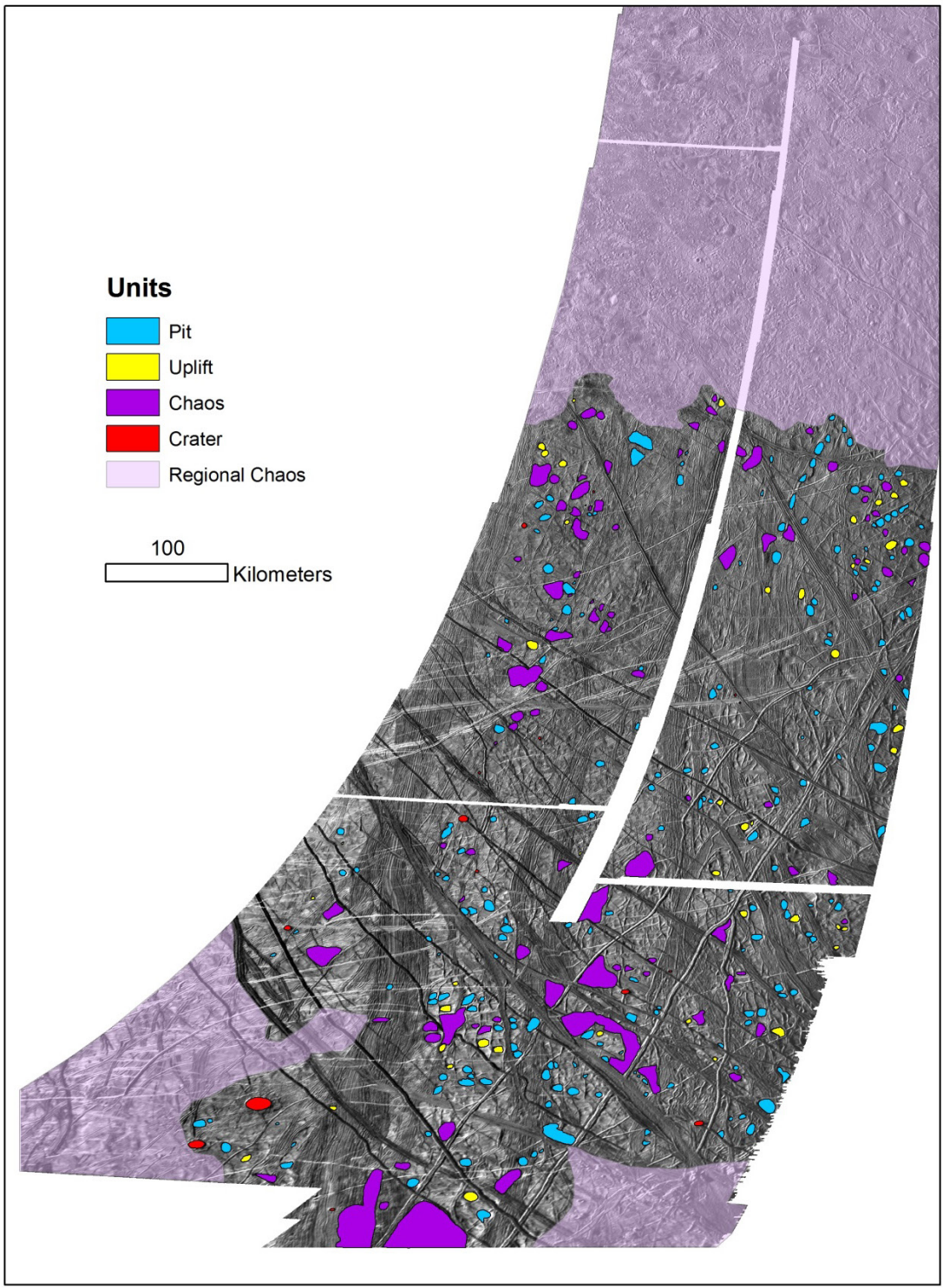

**Figure 2b: East RegMap – Southern leading region.** Scale bar applies to the latitude at which it is placed. Extent: -65° to -16°S and 70° to 107°W. Mapped area (not including the regional chaos) = $3.1 \times 10^5$ km$^2$.

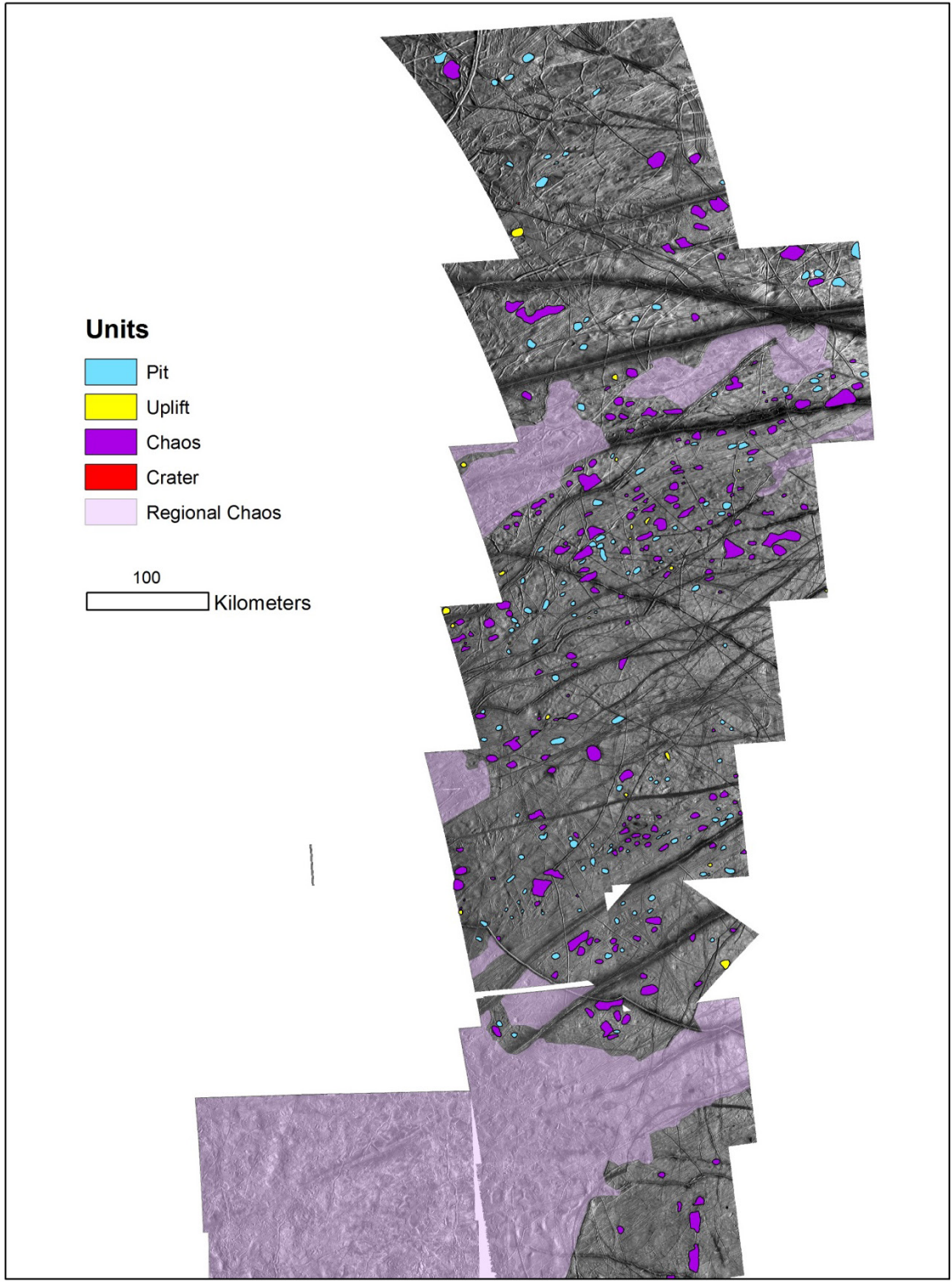

**Figure 2c: West RegMap – Northern trailing region.** Scale bar applies to the latitude at which it is placed. Extent: 0° to 62°N and 213° to 245°W. Mapped area (not including the regional chaos) = $4.3 \times 10^5$ km$^2$.

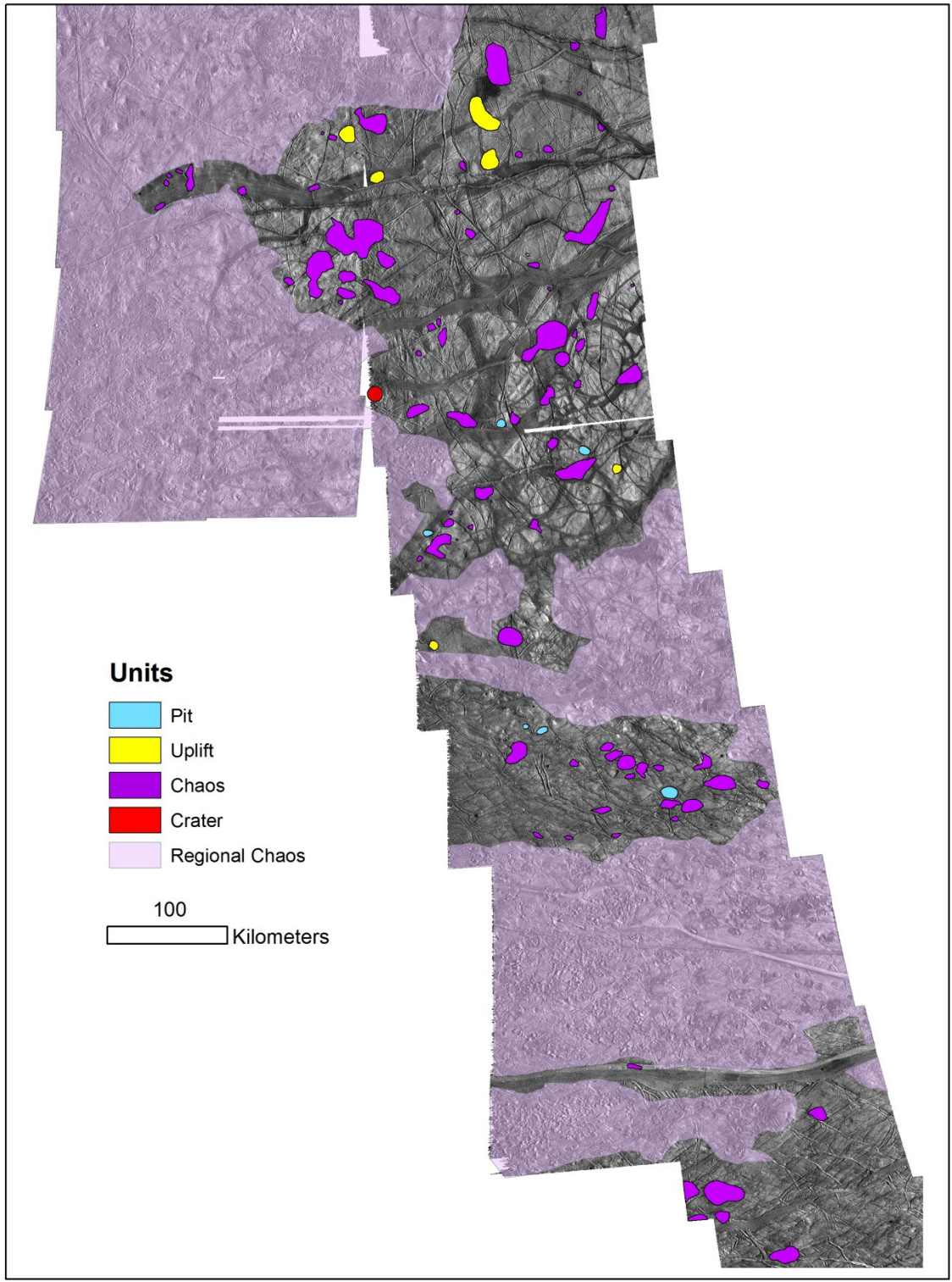

**Figure 2d: West RegMap – Southern trailing region.** Extent: -52° to -2°S and 207° to 245° W. Mapped area (not including the regional chaos) = $3.0 \times 10^5$ km$^2$.

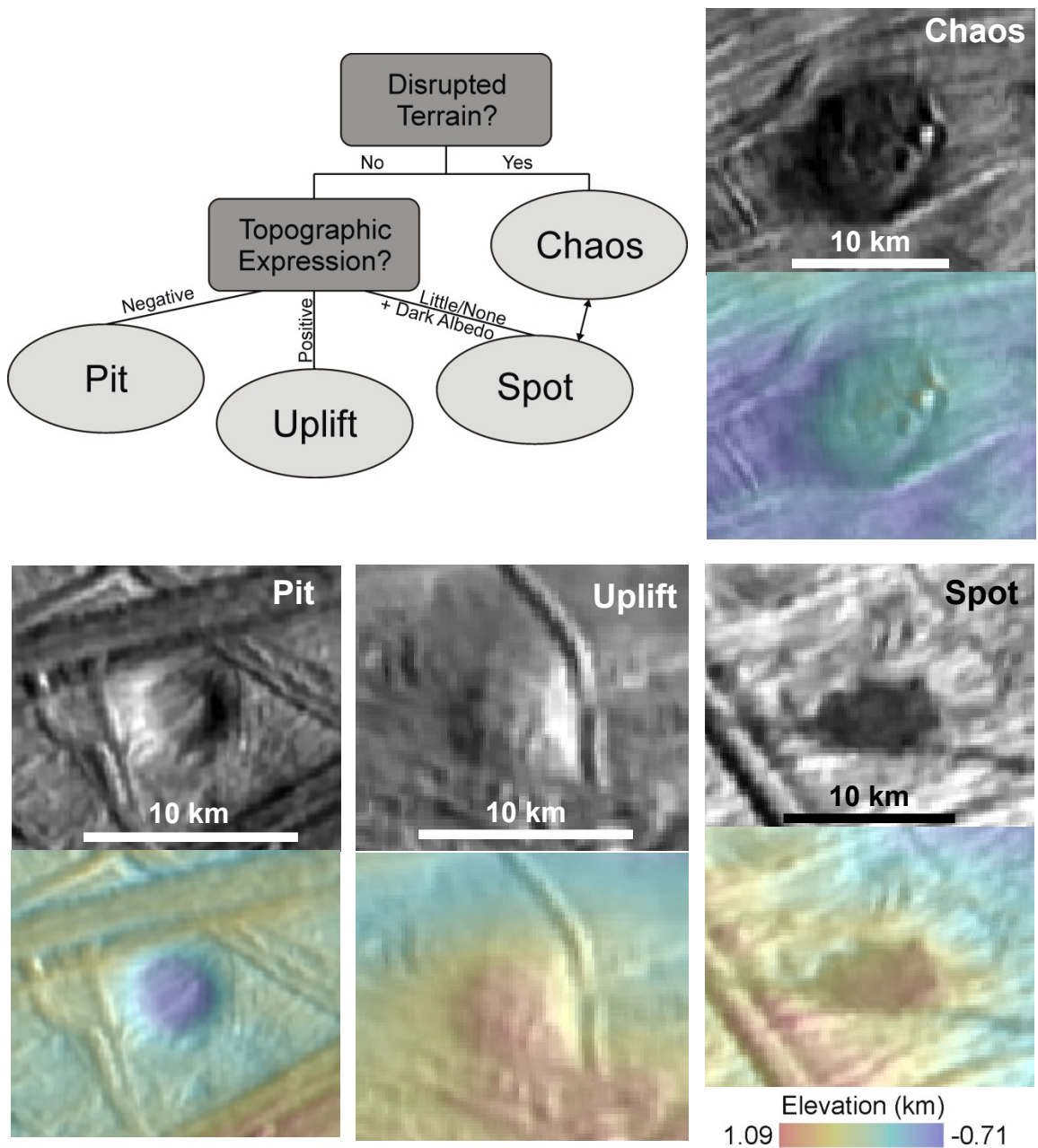

**Figure 3: Scheme used for designating feature classes and examples from the RegMaps.** Features with severely disrupted terrain (blocks or hummocky texture) were classified as chaos, while those with no or only slight surface disruption were further classified based on topography. Dark areas that might have been classified as spots in previous work were considered likely to be chaos and classified as such here. There are very few patches of the surface that appear to be smooth (neither ridged nor chaotic) at the resolution of the Galileo images.

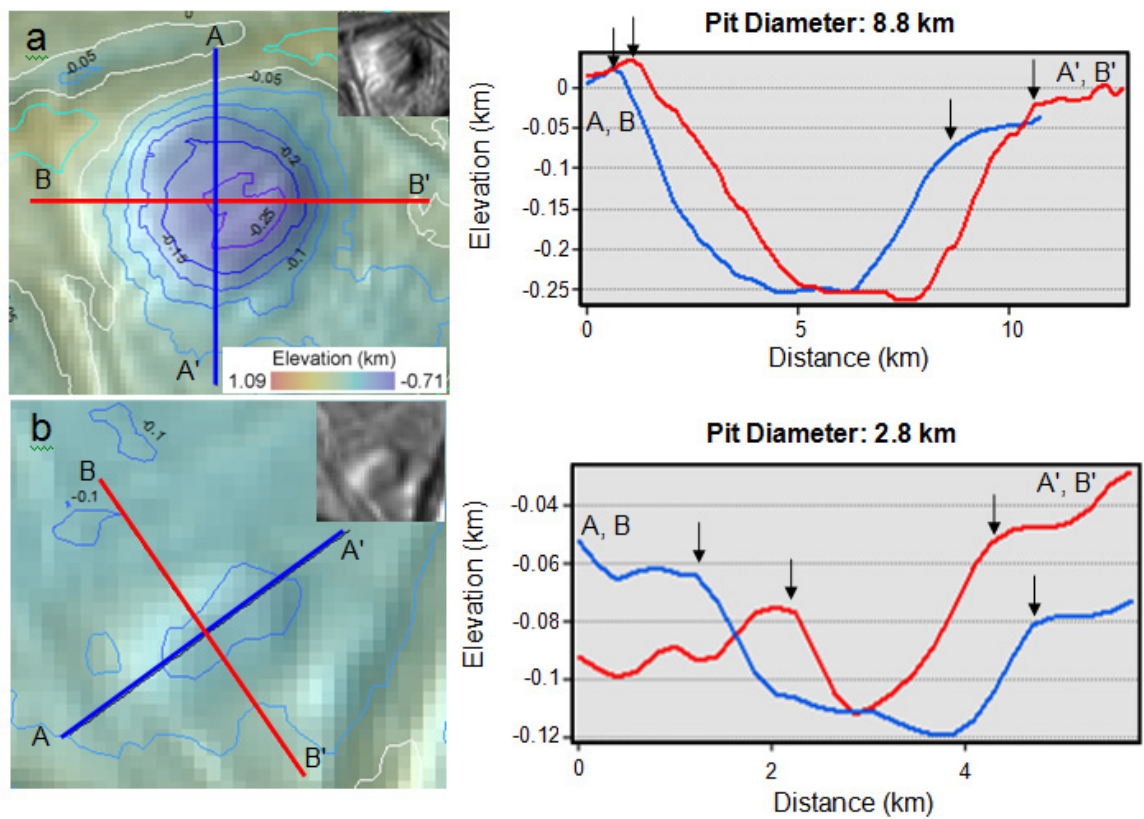

**Figure 4: Example pit profiles.** Shown for a) a large, circular pit (depth ~0.25 km), and b) a smaller, more elliptical pit (depth ~0.05 km). We map feature boundaries based on the inversion point in the topographic profiles (indicated by arrows).

## 2.3 Topographic Data

Topographic data were derived from albedo-controlled photoclinometry (PC) and crosschecked with stereo data where possible (see methods in Schenk, 2002; Schenk et al., 2004; Singer et al., 2012; Singer et al., 2018). Low-phase-angle images were used to model the surface albedo and significantly decrease systematic error in the topography derived from photoclinometry, but not all albedo effects can be removed. Due to the integrative nature of photoclinometry, errors in calculated slopes accumulate over large distances. This primarily affects large-scale variations in topography, for example, the topography on either side of a broad dark band may be artificially raised or lowered. For mapping of small features, such as those in this paper, this error does not accumulate substantially. The vertical error per pixel in the topographic data set can be derived from the photoclinometric uncertainty in the slopes, which are ~1-to-2 degrees per pixel for the regional maps. This equates to a vertical error of ~4-to-7 m per pixel at Regional Map resolutions. The systematic error in the depth/height of features given in the following section is more difficult to quantify as each feature is unique, and there may be small effects from both the error inherent in the photometric function or small-scale albedo effects. For features with more uniform photometric properties, random accumulating errors should be on the order of $\sqrt{n}$ × the pixel height error, where $n$ is the pixel scale of the feature of interest.

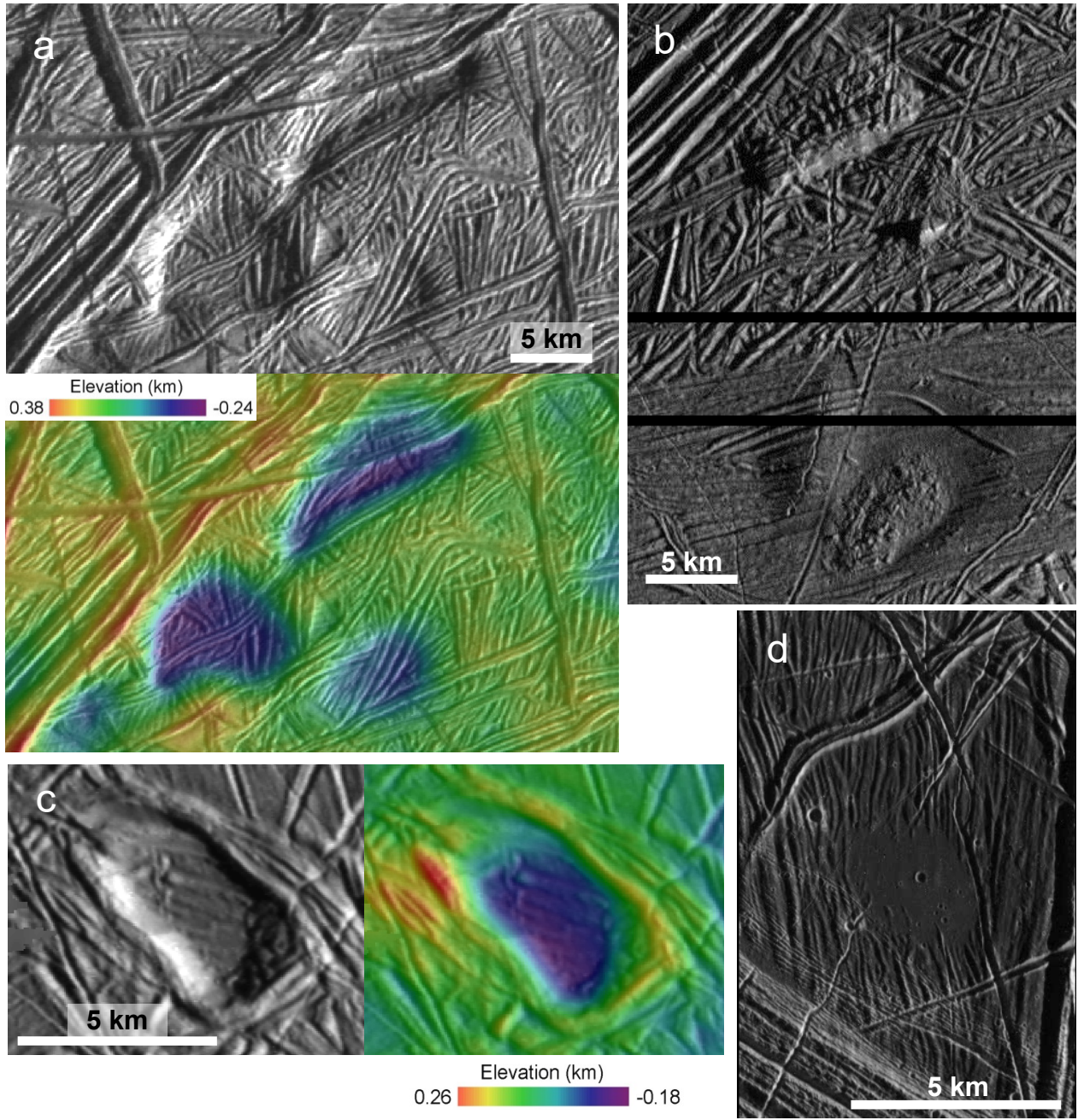

**Figure 5: Example feature as seen in high resolution images.** (a) Pits near Rhadamanthys Linea (65 m px$^{-1}$; also see full mosaic in Fig. B.16). (b) Uplift and chaos regions near Conamara Chaos (56 m px$^{-1}$). (c) Pit near Astypalaea Linea (41 m px$^{-1}$). (d) smooth, sub-circular surface patch near 6°N, 327°W (26 m px$^{-1}$).

## 3 Results

### *3.1 Size-frequency Distribution*

Feature areas were calculated geodesically (on a sphere) in ArcGIS, and their size is represented by the equivalent diameter (*D*) of a circle with the same area. A summary of the number and average diameters of mapped features is given in Table 1. For all feature types combined in the East RegMap we find effective diameters to be approximately log-normally distributed, with a peak between 4 and 7 km (Fig. 6a). For all features in the West RegMap effective diameters have a slightly broader peak from ~4 to 9 km (Fig. 6b). Looking at *pits alone*, both the East and West RegMaps have a peak (mode) near 5 km. There are fewer uplifts overall, but those on the East RegMap display a very broad peak around 3 to 8 km and those on the West RegMap have a peak near 5-to-7 km. The areal extent of collected regional and localized chaos mapped here is ~65% of the East RegMap and ~45% of the West RegMap.

A practical resolution limit for identifying approximately circular features (e.g., impact structures, pits) is 5 pixels across. This equates to a feature ~1.1 km in diameter for the RegMaps and much smaller for the high-resolution images. Although we did not restrict ourselves to this size when mapping, no features were found that could be confidently identified under this limit. Features smaller than 1 km in diameter can be seen, such as small dark pit-like features 3-to-5 pixels across, but these appear to be part of larger chaotic disturbances in almost all instances. It cannot be ruled out that smaller features exist below the resolution limits in the RegMaps, but what are mapped here are features that can be verified with topography and are plausibly single expressions of endogenic processes. Our study finds a peak in the size distribution for all pits, and indeed for all features combined at all resolutions, at ~5-to-6 km in diameter, well above the effective resolution for ~circular feature identification.

An examination of high-resolution images (~26-to-100 m px$^{-1}$) does not reveal any pits smaller than 3.3 km in diameter (Fig. 7). Ridges border two sides of this smallest pit, which is near Rhadamanthys Linea. Thus, it is possible the full expression of this pit is suppressed. The smallest feature overall seen in the high-resolution data is a 2.3-km-diameter subcircular chaos region, also near Rhadamanthys. In the RegMaps, the smallest pit mapped is 1.88 km in diameter, but is also bordered on two sides by ridges. The smallest pit not bordered on any sides by ridges in the RegMap is ~4 km in diameter.

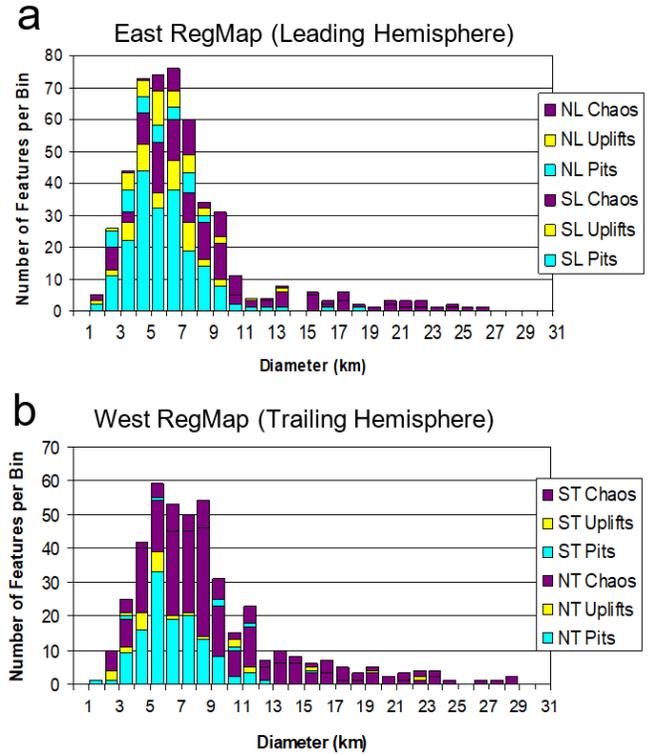

Figure 6: Histograms of effective feature diameters. Shown for (a) all features in the East RegMap and (b) all features in the West RegMap, also broken out by northern or southern hemisphere (e.g., NL stands for northern half of the leading hemisphere of Europa).

Table 1: Feature size summary.

| Size Statistics[1] | Pits | | Uplifts | | Chaos | |
|---|---|---|---|---|---|---|
| | East | West | East | West | East | West |
| Peak | 4–7 km | 5–6 km | 5–8 km | 4–6 km | 5–10 km | 6–9 km |
| Arithmetic Mean[2] | 5.2 ± 2.6 | 5.8 ± 3.4 | 5.7 ± 2.4 | 4.6 ± 1.9 | 10.1 ± 5.4 | 8.1 ± 4.8 |
| Median | 4.8 | 5.2 | 6.1 | 4.2 | 8.8 | 7.1 |
| Geometric Mean | 4.6 | 5.0 | 5.2 | 4.3 | 8.8 | 7.1 |
| Number | 231 | 133 | 83 | 28 | 158 | 277 |

[1]Statistics for diameters of features in km.
[2]Errors are one standard deviation.

Our results differ notably from those in Greenberg et al. (2003) and Riley et al. (2000). These authors argue that there is no characteristic spacing or modal dimension for pits, uplifts, or small-scale chaos, but rather they become increasingly frequent at smaller sizes/diameters, without limit. Greenberg et al. (2003) mapped pits and uplifts over the same RegMaps as our studies, and Riley at al. (2000) utilized the RegMaps as well as any areas imaged by *Galileo* where chaos could be identified, though neither study had the benefit of quantitative topography (shadows and shading on images can nevertheless be discriminated by eye). The reasons for these dramatically different conclusions are not clear, but we suggest three possible effects that could explain the discrepancy between these global surveys. First, Riley et al. (2000) and Greenberg et al. (2003) use area rather than length as a plotting statistic, which more strongly emphasizes the smallest-sized features at the expense of larger-sized features in what we find are closer to log-normal size-frequency distributions (see also discussion in Collins and Nimmo (2009) on this point). We remind the reader that the roll-offs in small-scale feature numbers that we find (Fig. 6) are well above resolution threshold for the RegMaps. Second, we confirm uplift and pit identification through their topographic expression in stereo-controlled-PC digital elevation models (DEMs), which as discussed below show a correlation of elevation or depression, as the case may be, with feature size. Hence the smallest uplifts and pits would have only slight topographic expressions, if any were recognized at all. Third, there may be an inherent difference in how depressions are identified and this may relate to resolution biases in severely lineated terrain of the type that dominate Europa. In images where ridges are no more than a few pixels across (as in the RegMap mosaics), sets of intersecting and truncated lineament sets can create the appearance of depressions that are in fact false or artificial basins formed by truncated lineament sets, especially on a planet where strike-slip offsets are so common (e.g., Kattenhorn and Hurford, 2009).

The dearth of true small depressions is borne out in virtually every high-resolution image of Europa (e.g., Fig. 7), and is further discussed in Appendix B, along with comparisons with other published studies. Nor would a continuation of increasing numbers of small-scale features to arbitrarily small sizes make much sense mechanistically. At some point ice shell thickness, or more precisely, ice lithosphere thickness must act as a throttle or delimiter on the surface expression of subsurface disturbances. The only way for arbitrarily small pits, uplifts and chaos to form would be if Europa's ice shell were arbitrarily thin, which we will argue below is not the case. Overall, we concur with observations by other workers (see Appendix B) that there are preferred or characteristic (i.e., modal) dimensions for ovoid depressions and uplifts, as well as small chaos, but further acknowledge that regional differences could also be present, pending the completion of a global map by some future Europa mission (such as *Europa Clipper*).

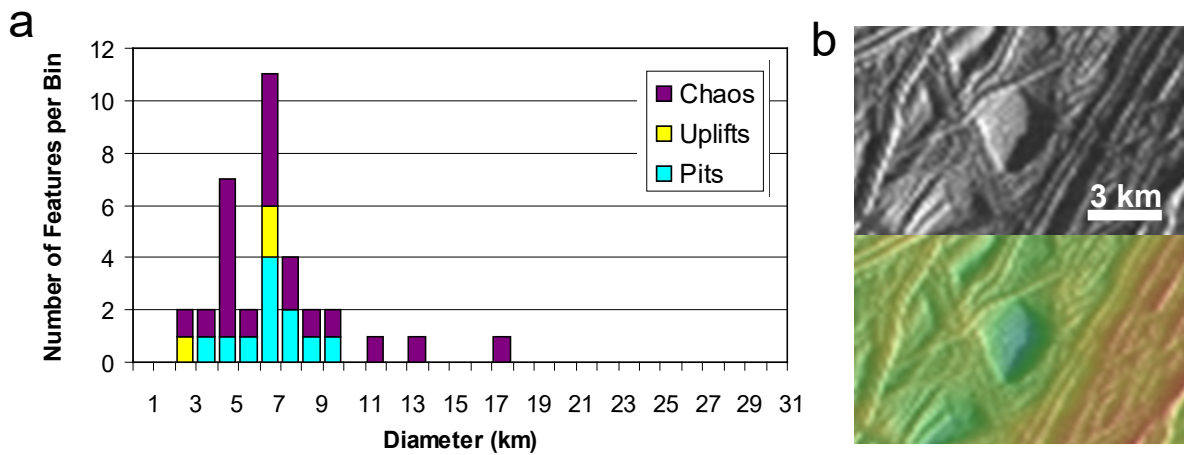

**Figure 7: High resolution mapping.** (a) Histogram of effective diameters for features mapped in high resolution images. (b) Smallest pit mapped in high resolution images, near Rhadamanthys Linea (65 m px$^{-1}$).

## 3.2 Feature Topography

The difference between the average elevation of the surrounding terrain (at the mapped feature outline) and the minimum elevation of a pit was used to find the depth ($d$). Similarly, the maximum elevation of an uplift was used to find the height ($h$). A summary of feature topography is presented in Table 2. The average pit depth is 0.12– 0.14 km but they range from very shallow (~10 m) to ~0.5 km deep (Fig. 8a). Uplifts are mostly ≤ 0.32 km high with an average of ~0.1 km (Fig. 8b), though some large uplifts in the southern part of the West RegMap are substantially higher (up to 0.8 km). Additionally, pit depth and uplift height both increase with increasing effective diameter (Fig. 9). Although there is some spread to the data, this general trend is consistent throughout the range of pit and uplift sizes from 1 km to over 20 km. The depth to diameter ratio ($d/D$) of pits ranges from 0.007 to 0.07, with an average of ~0.025—shallower than most impact craters (Schenk, 2002). The height to diameter ratio ($h/D$) for uplifts is similar to the $d/D$ for pits, it ranges from 0.007 to 0.06, with an average of ~0.025.

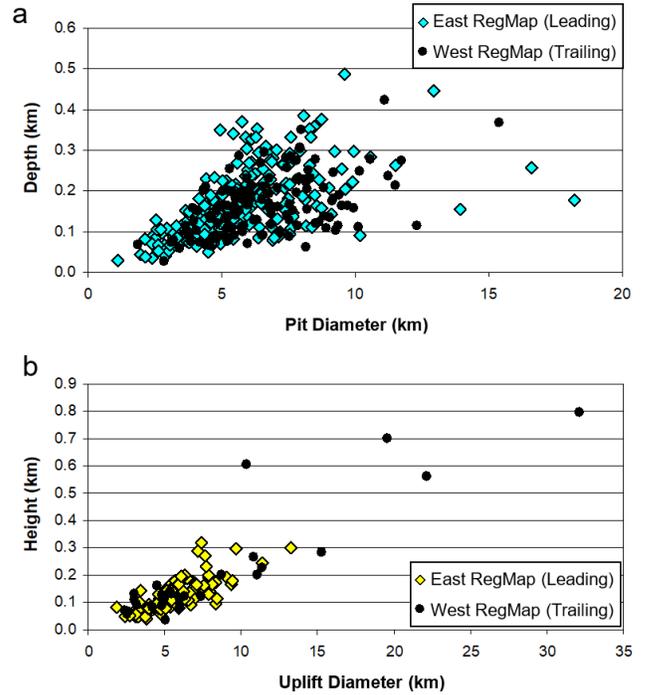

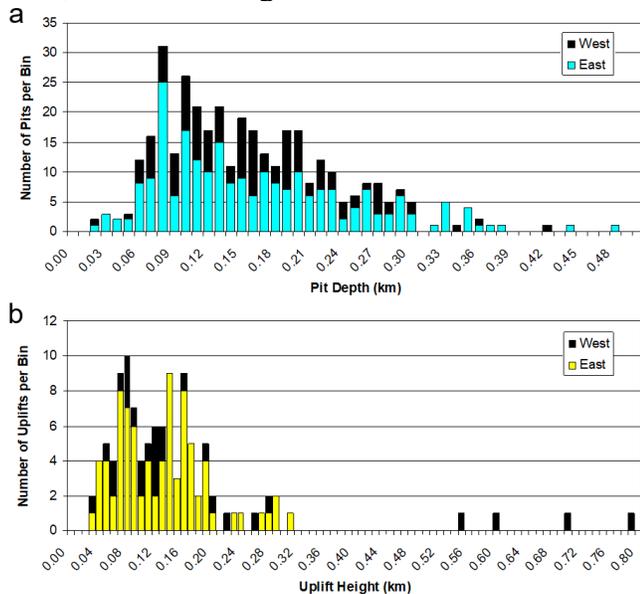

**Figure 8: Histograms of feature topographic signatures.** Shown for a) pit depths and b) uplift heights.

**Figure 9: Trends in feature topography with diameter.** Shown for a) pits, and b) uplifts. Both pits and uplifts show a general pattern for increasing topographic expression with increasing diameter.

## 3.3 Spatial Patterns

We tested for a latitudinal dependence of depth, diameter, or $d/D$ and found no strong patterns for mapped features on either the East or West RegMaps (Fig. 10). Similarly, there were no spatial patterns found for uplift height or $h/D$. The only spatial correlation seen for pit depths was that, in a few cases, several pits trending along the same line had similar depths (Fig. 11). The features studied here are generally found in the ridged plains where large-scale chaotic disruption has not occurred, but they are not uniformly distributed in the ridged plains. Previous studies attempted to characterize the average spacing between chaos features as a way to infer the ice shell thickness assuming convective upwelling as the cause (Pappalardo et al., 1998; Spaun, 2002). In our mapping areas,

Table 2: Feature topography summary.

| Topographic Statistics | Pit Depth (km) | | Uplift Height (km) | |
|---|---|---|---|---|
| | East | West | East | West |
| Range | ~0.02–0.5 | ~0.02–0.43 | ~0.03–0.32 | ~0.03–0.8 |
| Arithmetic Mean[1] | 0.14 ± 0.09 | 0.12 ± 0.07 | 0.13 ± 0.06 | 0.09 ± 0.04 |
| Geometric Mean | 0.11 | 0.10 | 0.10 | 0.09 |

[1]Errors are one standard deviation.

a) East RegMap - Leading

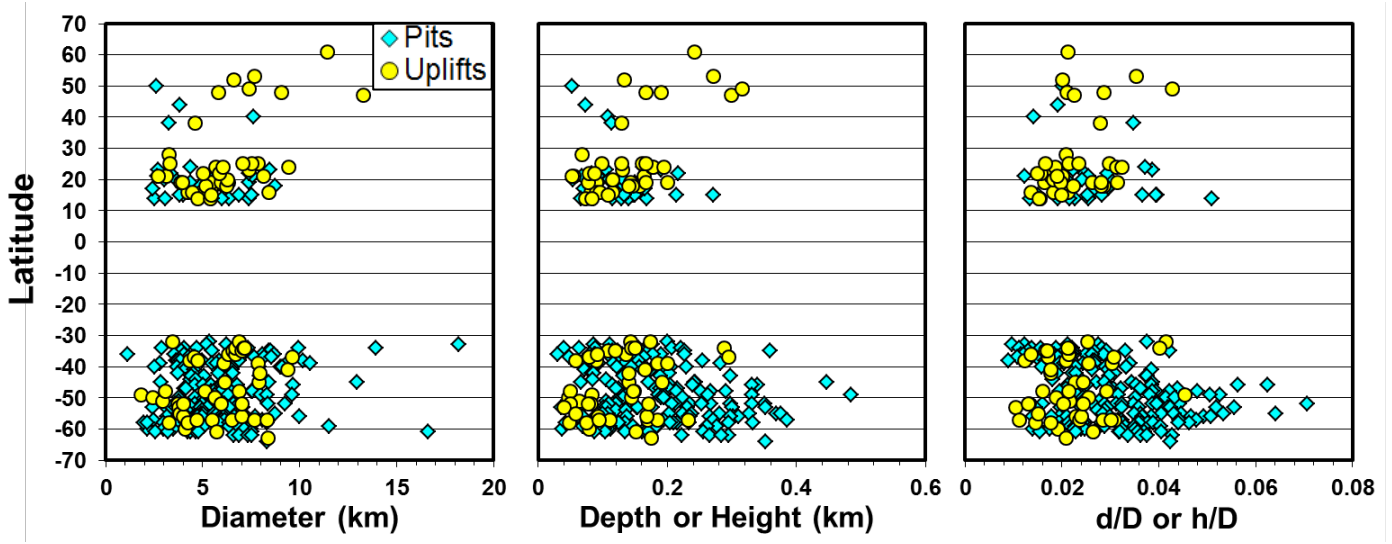

b) West RegMap - Trailing

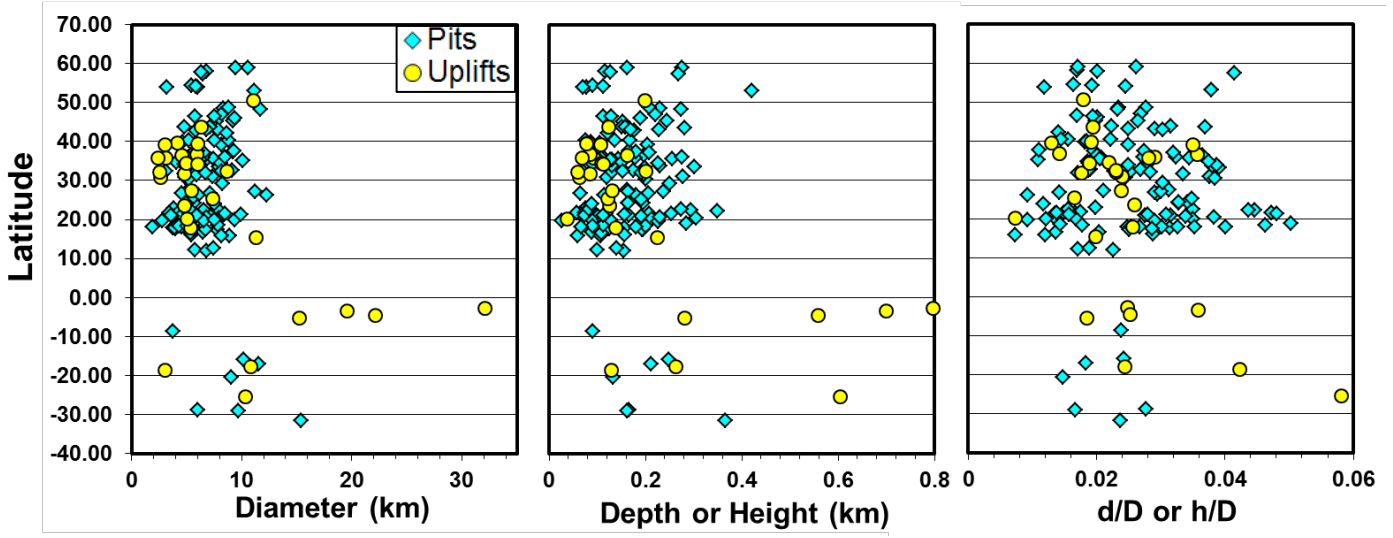

**Figure 10: Latitudinal dependence of feature size or topographic expression.** No strong trends were found for feature characteristics with latitude, although some regions have more features than others as described in the text. Here we consider feature equivalent diameters, pit depth, uplift height, pit depth-to-diameter ratio ($d/D$), and uplift height-to-diameter ratio ($h/D$).

features (of all types) were sometimes clustered locally into patches, but spacing between features within a given patch was not regular, and would depend heavily on where one arbitrarily defined the patch/regional boundaries. Thus, we did not pursue this topic further here. As is easily seen by eye, features in the same area sometimes have similar color or albedo expressions, but often they vary, and no specific spatial patterns were identified for these aspects either.

## 4 Discussion

### *4.1 Feature Formation Mechanism*

The formation of pits, uplifts, and small chaos features are broadly considered to be related (e.g., Collins and Nimmo, 2009; Noviello et al., 2019, and references therein). These disturbances are clearly endogenic in origin, and over the years several models or mechanisms have been proposed or elaborated as explanations. We address the principal ones among these—diapirism, intra-crustal sill formation, and melt-through—in the light of our results.

Diapirism, or more generally, solid-state convection, if it occurs, can stress or deform Europa's icy surface, even in the stagnant lid regime in which the outermost cold layer of ice is immobile. For vertical or normal stresses to create surface depressions (or pits) of the depths seen at Europa (in excess of 100 m) requires quite specific conditions, however (Showman and Han, 2004). If the viscosity ratio between the cold surface ice and warm basal ice (i.e., that in contact with Europa's ocean) is too low, then convective stresses are low overall and the surface deformation minimal; in contrast, when this viscosity ratio is high (as implied by the strong temperature contrast across the shell), then the stagnant lid acts as a mechanical filter and strongly limits surface deflection (see Fig. 7 in Showman and Han, 2004). Even for the very limited range of viscosity ratios for which sufficiently deep surface depressions can form, and for which there is no a priori justification, the pits that do form are much broader (≳10-km across) than found on Europa. Moreover, the calculations in Showman and Han (2004) are 2D Cartesian, and do not address the likely full 3D planform. For a floating ice shell, with basal free-slip, the 3D planform is likely one of central upwellings and peripheral, sheet-like downwellings, not isolated downwelling plumes or drips (e.g., Trompert and Hansen, 1998)[1]. Incorporating a plastic yield limit to the stagnant lid or ice lithosphere does not improve matters either (Showman and Han, 2005). Hence, we reject "normal" convective overturn within a pure ice shell as likely mechanism for the production of pits, uplifts, or small chaos.

If, however, the ice shell is compositionally stratified (salty ice over clean ice, such as might result from cooling and thickening of the ice shell) then the strong compositional buoyancy of the initial convective flow could lead to surface deformations of the required (observed) amplitude, albeit on generally broader horizontal scales than observed on Europa, and then only if the vertical viscosity contrast (top-to-bottom) across the shell is not too great (Han and Showman, 2005).

Now, there is a preferential feature size for (at ~5-6 km in diameter), as well as a broad range of diameters seen for the pits, uplifts, and small chaos features (Fig. 6). *These* distributions are consistent with the predictions of diapirism in a general sense, in that at any given location the ice lithosphere acts as a mechanical filter. Internal heterogeneities less than a characteristic deformation scale (e.g., flexural radius, pre-existing fault spacing) will be biased towards features of those scales. In addition, the correlation of pit sizes and depths (Fig. 9a) is more easily understood within the context of diapirs occurring in the shell along with other processes. Continued tidal heating in an ascending icy diapir could potentially keep it warm (due to enhanced tidal dissipation in the warmer ice of the diapir compared with its surroundings) and thus promote a density contrast to drive the continued rise of the diapir (e.g., McKinnon, 1999). This process could potentially induce partial melting (Sotin et al., 2002; Mitri and Showman, 2008), or if the ascending diapir encounters saltier ice, induce eutectic melting of that ice (Schmidt et al., 2011). The (possibly salty) water released could drain back into the ocean below, or if such heating occurs when the diapir is near the surface of the icy lithosphere, could form a water-rich lens (Sotin et al., 2009 and references therein). This loss of volume through melting could lead to isostatic compensation in the form of a pit. In this scenario, the larger the diameter of the diapir the more volume is available for potential melting and volume loss, resulting in a deeper pit.

---

[1]The aligned pits in Fig. 11 are intriguing in this regard, though this is not characteristic of the geometric arrangement of pits so far mapped on Europa.

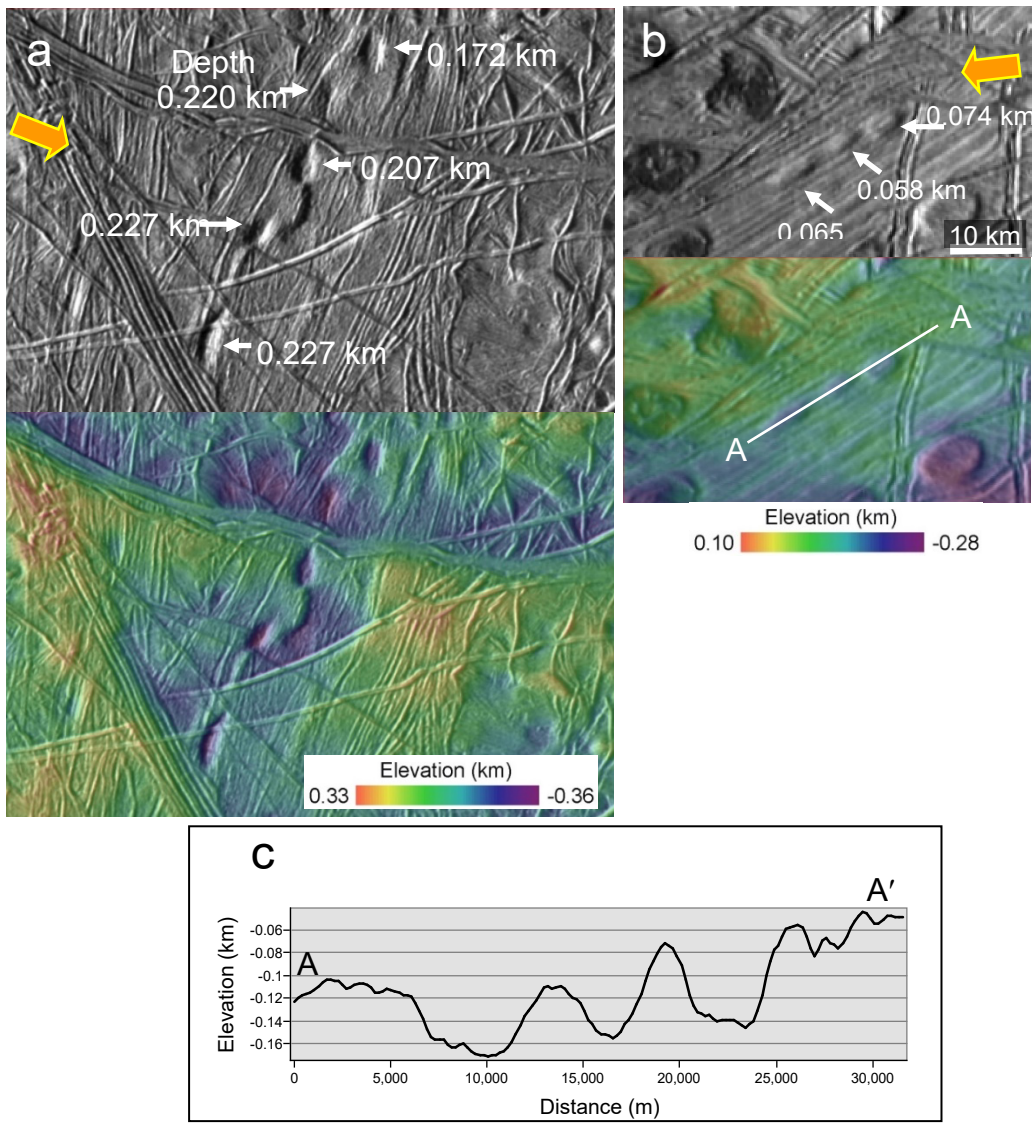

**Figure 11: Linearly arranged pits.** Images and topography in (a) the East RegMap (depths shown for 5 pits near -35°S and 76°W), and (b) the West RegMap (depths shown for 3 pits near 22°N and 223°W). Wide orange/yellow arrow indicates approximate direction of incoming sunlight. (c) Topographic profile over linear pits in West RegMap.

More recent models have focused on emplacement of water-filled sills at relatively shallow depths with Europa's crust (Schmidt et al., 2011; Michaut and Manga, 2014; Manga and Michaut, 2017; Craft et al., 2019). The presumed source of this water is the (possibly pressurized) ocean. Ocean water would rise in vertical cracks to a position of neutral buoyancy, or perhaps more likely, a rheological boundary such as the brittle-ductile transition (BDT)[2], where it would then spread laterally. An intracrustal sill origin for Europa's pits presents some issues, however. Foremost is that the intrusion of a lens of new material bulks up the ice shell, which is not an obvious path to generating substantial surface depressions. Following the detailed modeling in Michaut and Manga (2014) and Manga and Michaut (2017), we find the following: 1) A fluid-filled, horizontal crack or sill that is not inflated is inadequate on its own to create pits (or uplifts) of the vertical scale observed, even assuming a much reduced elastic Young's modulus for the surrounding ice; 2) the elastic ice either above or below the sill must be thin and/or weak enough in order for the sill to inflate vertically to form a laccolith or lopolith (and thus yield the observed topographic signal); and 3) the diameter ($D$) of the saucer-shaped sill can be related to its depth ($d$) by $D \approx 4.8d$, because the margins of such sill cracks bend sharply upward at that limit. This suggests, if an intracrustal sill origin is viable, that feature scales of ~5-6 km imply water lenses (either extant or former) at ~1 km depth. The consistency of pit shape with this model is discussed below.

Pits, as opposed to uplifts, require rather thin elastic layer support beneath nominal, relatively dense water lenses (Manga and Michaut, 2017). This is consistent with sill/lopolith formation at or near the BDT. The strains implied in such ice ($\gtrsim$1%) exceed any plausible yield strength for ice, however, and substantial heat will diffuse into the subjacent ice before the sill freezes as well, so in reality such a lopolith will likely evolve into a descending water diapir (Michaut and Manga, 2014). Such a descending mass should generate much higher viscous stresses than (e.g.) those in the numerical study of Showman and Han (2004), so coupled with the thin ice shell above the initial lopolith, sufficiently deep pits are plausible (though this remains to be modeled quantitatively). Pit depth should scale with lopolith/diapir size, consistent with Fig. 9a.

In contrast to the above, if water nearly melts through to the surface to form a pit, as in the melt-through model (O'Brien et al., 2002; Greenberg et al., 2003), then the floors of the pits would likely initially reach a uniform level of isostatic compensation based on the shell thickness, not an increasing depth with increasing pit size. This does not mean that melt-through does not occur, only that most pits and other small features surveyed here probably did not form exclusively in this manner. Our results show that no small pits are as deep as their larger counterparts, which implies pits did not all start out at an initially deep, isostatic, level. Viscous relaxation of topography could make pits shallower over time, but larger pits would relax faster, and some small pits should still be fairly deep. Relaxation may account for some of the spread in feature depths/heights for a given diameter. These are, however, relatively small wavelength features, and presumably relatively geologically young because they are not disrupted by later ridge formation. The entire surface of Europa is deemed to be only 40-90 million years old (~20-200 Ma for the entire spread of the error bars) based on crater spatial densities (Bierhaus et al., 2009), which leaves limited time for pits/uplifts to relax. Models of the relaxation of the fold-like features seen at Astypalaea Linea with a wavelength of ~25 km are dependent on the surface temperature and thus latitude (Dombard and McKinnon, 2006). For timescales of 10 to 100 m.y., considerable relaxation is only seen models of equatorial features on Europa, where the surface temperature is higher. Pits may also be shallower if elastic support inhibits full isostatic compensation of any subsurface volume loss.

The close proximity of some features to each other is also potentially consistent with diapirism, as it is conceptually simpler to maintain the original terrain between the features with diapirism. Melt-through would presumably lead to a very thin shell between two nearly-adjacent features, and thus it might be difficult to preserve

---

[2]Horizontal bedding planes are probably uncommon on Europa, as it is a world whose stratigraphy is likely predominantly vertical (Lucchitta and Soderblom, 1982).

the pre-existing terrain between the features with this mechanism.

Pit morphology varies, but one feature of note is that many pits have a relatively sharp change in slope, rather than a gradual sloping transition, from the pit walls to the surrounding terrain or the pit floors. This gives the features more of a "pie-pan" shape, as opposed to a bowl shape. Near Rhadamanthys Linea, the pit walls retain pre-existing ridges, and although steep, are somewhat rounded (Fig. 5a, Appendix B). The pit seen at high resolution near Astypalaea Linea has especially sharp, vertical walls (Fig. 5c). If subsurface melt occurred deep in the shell, one would expect a longer wavelength surface response typical of a bending elastic shell. The more sharply-defined, localized expression of many pits may indicate that any subsurface volume loss (melting by any mechanism) is occurring sufficiently close to the surface to cause pit formation. Thus if pits formation is initiated by diapirism in the shell (e.g., Schmidt et al., 2011), the diapirs must be ascending to near surface levels. In the context of the sill model of Manga and Michaut (2017), upward rotation of the sill geometry to form a saucer shape, due to the interaction of stresses at the crack tip and the bending stresses of the overlying ice, is predicted (as noted above)[3]. This effect of the highest tensile stresses occurring along the sill edges and inducing fracturing is also seen in numerical models (Craft et al., 2019). Failure and collapse along these fractures would be consistent with steep pit walls observed. Future modeling should be able to predict the depth to melting by both matching the depth of pits (as noted above for intracrustal sills) *as well as* the slope of the walls. Near-surface water has clear astrobiological implications, and influences predictions for what might be observed with REASON (the ice-penetrating radar on *Europa Clipper*).

*Uplifts:*

Several concepts for forming uplifts have been proposed, including diapirs pushing up the surface (e.g., Pappalardo and Barr, 2004; Schenk and Pappalardo, 2004) and injection and re-freezing of sub-surface water (Michaut and Manga, 2014; Manga and Michaut, 2017; Craft et al., 2019). These mechanisms apply to uplifts as defined and measured here, where the pre-existing surface is retained and there is no strong indication of emplacement of new material onto the surface, as might occur in a cryovolcanic context (Quick et al., 2017). The data presented here do not discriminate between these mechanisms. From a mechanics point-of-view, however, a purely diapiric mechanism for uplifts requires that the ascending plume reach, if not breach, the stagnant lid and ice lithosphere. This may be more readily accomplished if the source of buoyancy is compositional rather than thermal, as is the case of a pure ice diapir ascending through a (somewhat) salty ice shell (Pappalardo and Barr, 2004; Han and Showman, 2005). A saucer-shaped, water-filled sill is also a plausible explanation for uplifts, as the addition of bulk to an ice shell naturally favors uplift of the surface. In the model of Manga and Michaut (2017), uplift is the natural outcome (as opposed to depression) as long as the sill forms reasonably close to the surface but well within the elastic lithosphere, so that basal elastic support for the laccolith (or lopolith) is available. For similar sill geometries and depths, uplift is a favored outcome for thicker ice shells, i.e., those with thicker ice lithospheres (lower heat flows). In contrast, thinner ice shells (higher heat flows) tend to favor depression or pit formation.

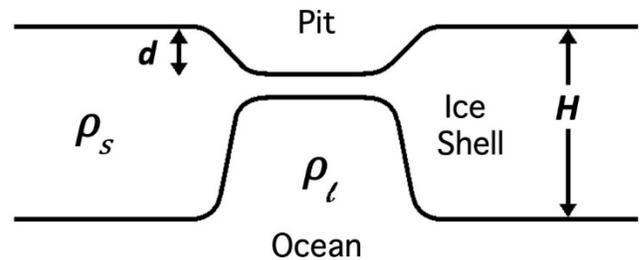

**Figure 12: Cartoon of pit isostasy.** Parameters for estimation of *minimum* ice shell thickness ($H$) from isostasy using pit depths ($d$), and various estimates of ice shell densities ($\rho_s$) and ocean densities ($\rho_l$). True shell thickness estimate should be considerably larger than this minimum value. Note that the pit may, or may not, be wider than the ice shell is deep.

---

[3]For terrestrial and experimental examples, see Polteau et al. (2008) and Galland et al. (2009).

The trend of increasing uplift height with increasing size and the similar size of uplifts and pits gives further indication that the subsurface processes creating them are related. Both large and small chaos regions have a variety of topographic and planform expressions. Some are elevated with respect to the surround terrain, some undulate (have both low and high areas), some are at the same level or depressed with respect to the surrounding terrain, some are dome-like, and many have more irregular shapes. Perhaps chaotic terrain is created by a more energetic version of the same processes that creates pits and uplifts, or the processes occur at shallower depths.

### 4.2 Shell Thickness

On Earth, the formation of large quasi-circular, ellipsoidal, or ameoboid depressions is usually attributed to withdrawal of liquid material from below, in the absence of surface processes. Examples include magma withdrawal and caldera formation, dissolution of limestone by groundwater and formation of karst depressions, dissolution of underground salt (halite) and formation of sinkholes, and melting and drainage of water ice in permafrost terrains to form thermokarst. On an icy satellite such as Europa, volume loss may also be associated directly with melting of water ice or water-ice rich terrains, due to the large negative volume change upon melting of water-ice, which could be amplified by drainage to the ocean below.

The apparent continuum of pits, uplifts, and small chaos features strongly suggests that thermal anomalies (whether diapirs, water-filled sills or localized melt-through) underlie all these features, or did at the time of their formation. As such, ice beneath pits is expected to be warmer and weaker than average, and arguably able to better respond to vertical buoyancy forces so as to reach or at least approach a state of isostatic balance. Pits, with their minimal surface disruption, are plausibly caused by such subsurface loss of ice shell volume due to melting. Regardless of formation mechanism, we can estimate a lower limit for shell thickness using a simple Airy isostasy model. In a theoretical endmember case where the column of material below a pit has been replaced by liquid except for an arbitrarily thin lid (cartoon in Fig. 12), the shell thickness $H$ is given by

$$H = \rho_l/\Delta\rho \; d, \quad (1)$$

where $\rho_l$ is the density of the liquid, $d$ is pit depth, and $\Delta\rho$ is the density contrast between the denser liquid and the shell ($\rho_l - \rho_s$). A plot of predicted shell thickness is given in Fig. 13 for several possible ice and ocean compositions (described in greater detail in Appendix C). The eutectic ice-meridianiite-mirabilite ocean compositions are almost certainly extreme (too sulfate rich; McKinnon and Zolensky, 2003), so the range of likely minimum ice shell thickness is more restricted. A plausible range of $\Delta\rho$ and conservative maximum pit depths of ~0.3-to-0.5 km imply *minimum* shell thicknesses on Europa of ~3-to-8 km.

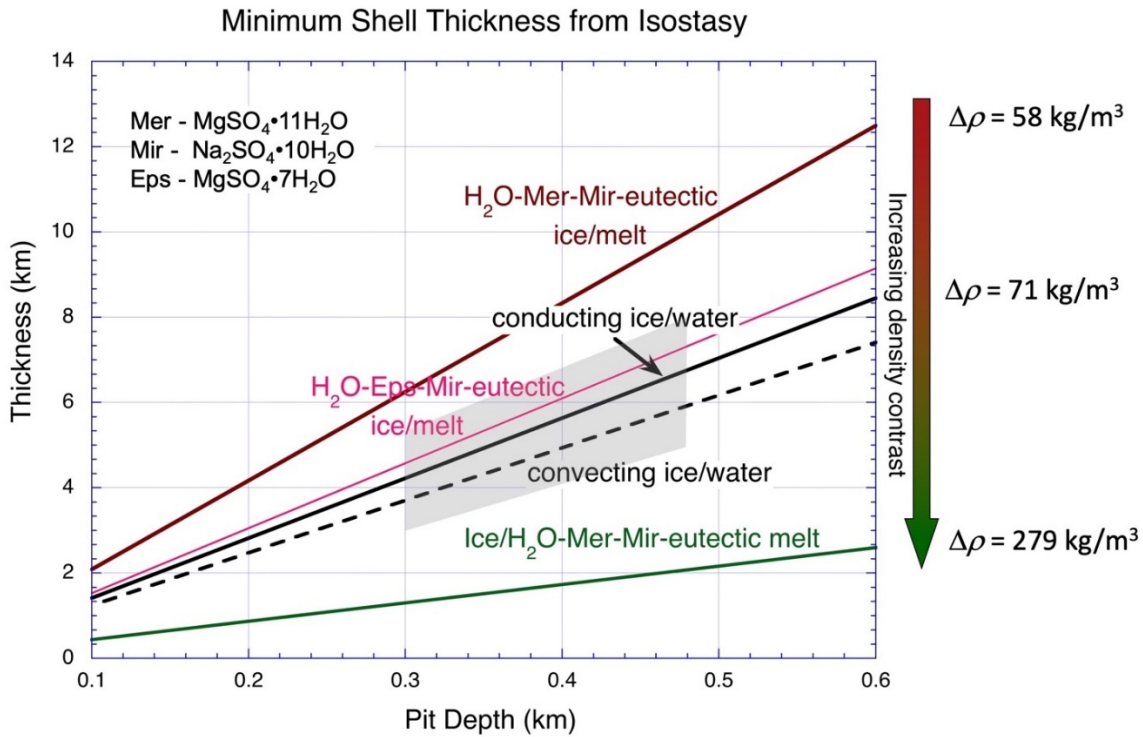

**Figure 13: Minimum ice shell thicknesses from isostasy.** Possible ice shell/ocean compositions are discussed in Appendix C. Using the deepest 5% of the pits (~0.3-to-0.5 km deep), and less extreme compositions leads to *minimum* ice shell thickness predictions of ~3-to-8 km (grey parallelogram).

A cross-section through a europan pit would not necessarily closely resemble Fig. 12, although it could in the melt-through model (Fig. 16b in Greenberg et al. 1999, O'Brien et al. 2002). As long as isostasy is maintained, the denser liquid that underlies the pit can be placed anywhere in the vertical column under the pit, such as near the base of the ice shell or distributed within the ice column; the latter might pertain to tidally or compositionally driven melting in an ice diapir. The point is that ostensibly more realistic configurations would imply thicker shells. The addition of ice to the column under the pit will only increase the total ice shell thickness estimate. We note that many pit depths easily exceed the ~40-to-250 m raft/block heights in Conamara Chaos (Williams and Greeley, 1998; Schenk and Pappalardo, 2004; Skjetne et al., 2021) so our minimum shell thickness estimates are corresponding larger than those calculated for the Conamara rafts from isostasy by Williams and Greeley (1998)[4]. At least one region of chaotic blocks outside of Conamara does exhibit some taller blocks (that do not appear to be tilted) with heights up to ~500 m (Skjetne et al., 2021).

The shell thickness calculated here applies to the time of feature formation, and it assumes isostatic balance was reached and that the topography has not since relaxed. If some or all of these features have viscously relaxed since their formation, this again implies that 3-to-8 km is a minimum, as the pits would have been deeper originally.

As intriguing as these shell thickness limits are, they do not easily inform intra-crustal sill models. Saucer-shaped *intrusions* of ocean water into Europa's mid or upper ice shell do not, strictly speaking, affect potential isostatic balance at all. In order for the surface to deform, either up or down, there must be either elastic or dynamic, viscous support, and this support must be maintained (Michaut and Manga, 2014; Manga and Michaut, 2017).

## 5 Conclusions

A peak in the size-frequency distribution for all feature types is seen at ~5-to-6 km in diameter. This is presumably representative of the thermal perturbation or anomaly below. A survey of limited high-resolution images (~27-to-185 m px$^{-1}$) does not reveal any pits smaller than 4 km in diameter (that are not bordered by ridges). Most pits are ~100 m deep, but a few are as deep as ~500 m. This observed range of pit depths fits the predicted depths from some numerical simulations of convection in Europa's shell (Showman and Han, 2004; Han and Showman, 2005), but pits observed in our study are generally both deeper and, more importantly, less wide than those achieved in the simulations, and observed pits have a wide range of morphologies. Larger pits are consistently deeper than smaller ones. Future models may reveal that more than one process can produce the observed trends of depth/height with feature size, but the results of this work favor (in the sense of being most consistent with) initiation of pit formation by a rising diapir and tidal or compositional melting, or lopolith formation by injected ocean water, as opposed to pure melt-through, for the majority of features. Pure melt-through would predict pits should all have a fairly uniform depth, regardless of scale. Subsequent processes occurring after melt is generated (e.g., brine mobilization, cryovolcanic or cryomagmatic processes, re-freezing) may result in the vast variety of features observed on Europa's surface.

---

[4]Sparse shadow-length measurements in Williams and Greeley (1998) indicated block tops only ~50-150 m above the surrounding matrix. Unfortunately, they had no control on the "brash" ice content of the matrix when elevation differences were frozen in, so the assumption of pure liquid densities in their Fig. 3 underestimates raft and block thicknesses.

A conservative estimate based on some of the deepest pits suggests a *minimum* shell thickness of ~3-to-8 km at the time and location of feature formation, if they formed in isostatic balance. Such lower limits to ice shell thickness do not in and of themselves settle the so-called debate between thin and thick shells (McKinnon et al., 2009), because in reality any shell thickness less than 10 km—for most researchers—probably counts as "thin," though 10 km (e.g.) is likely not thin enough for some (see Greenberg and Geissler, 2002). That Enceladus' floating ice shell turned out to be as thin as it has (~20 km; see Hemingway et al., 2018), is testament to Nature's ingenuity, and a cautionary tale. Our estimates might also appear to favor a thicker ice shell, as indeed any potential evidence for (or consistency with) diapirism does, but Europa is a dynamically evolving (Lainey et al., 2009) and likely thermally evolving world (e.g., McKinnon et al., 2009). Substantial ice shell thickness variations over time are plausible, and our conclusions regarding formation mechanism and shell thickness strictly refer to the epoch of feature formation in these regions, conclusions which are testable (e.g., by radar sounding and profiling, high-resolution imaging, and compositional measurements) by future missions such *JUICE* and *Europa Clipper*.

## Acknowledgments
We thank the NASA Planetary Geology and Geophysics Program (grant NNX11AP16G to WBM) and the NASA Earth and Space Science Fellowship (KNS) for funding this work. KNS also thanks Dr. Bob Pappalardo and Dr. Geoff Collins for discussions about an early version of this work, and Teresa Wong for discussion about convection in Europa's shell. The authors also thank helpful reviewer comments that improved this manuscript.

## Appendix A: Extended Discussion of Mapping Method

Previous studies have considered the effects of resolution and lighting geometry on the mapping of chaos (Hoppa et al., 2001; Neish et al., 2011). Chaos is most often identified based on its hummocky texture, and thus some of these effects will not be the same when mapping a feature defined by its topography instead (pits and uplifts as mapped in this study). The texture of chaos varies in scale, but individual blocks may be below the resolution limit, and in some cases chaos is primarily hummocky "matrix" consisting of finer scale blocks/broken-up ice. Resolution clearly affects mapping, and while it is generally true that one will be able to see smaller features with a better resolution, the effects on mapping chaos features, which are all unique – as opposed to features like craters, which have a fairly uniform and predictable shape – can be convoluted. In some cases, small, individual chaos features near each other may even be revealed to be part of one larger feature at high resolution (as the finer-scale fractures between the features were not visible at lower resolution). Thus, it is possible, but not always the case, that smaller features will be mapped at higher resolution.

We emphasize that mapping of chaos based on texture was not the main goal of this paper. We focused on topographic features (pits and uplifts) and sub-circular chaos with relatively distinct boundaries. As mentioned in the body of the paper, it is possible that small pits may exist below the resolution limit, but we do not observe any tiny pits in the high-resolution images. Also, based on the possible formation mechanisms, it would be difficult to form very small pits (hundreds of meters in diameter or smaller). Melt-through would require the heat source to "turn off" right as the shell was melted through to produce only a tiny pit (assuming the heat source was effectively a point source to begin with, and the shell was very thin). Diapirs on the scale of 100s of meters may exist in the shell, but in some size limit it must be difficult for a feature to exhibit a surface expression; the upper brittle lithosphere would presumably be strong enough to resist subsidence for the small stresses associated with small volume changes.

Although every attempt was made in this work to be internally consistent and use topography to verify features, mapping inherently involves decisions by the investigators. To see if a different definition of features would lead to different results, we also mapped using two other definitions: one more conservative and one less conservative than that in the main paper. For the category definitions below, four "sides" of a feature are defined by taking the circumference of a feature and dividing it by four (approximated by eye).

*Least conservative feature definition:*
- Pits and uplifts were defined as anything that could be construed as a feature and had some topographic signature, but not including features bounded on all sides by ridges (as in, the topographic low is only an apparent one because the ridges surrounding it are higher than the average terrain).
- In a few cases features in this category are bounded on 3 "sides" by ridges, but mostly only 0-2 sides (out of the 4 total).
- The least conservative definition of chaos involved mapping anything that could be construed as a finite, quasi-circular patch of chaos, primarily based on the observation of chaos texture (broken up blocks or hummocky matrix) but also including some dark albedo patches with possible chaos texture.

*Moderately conservative definition (used in paper):*
- Features mapped were bounded on less than or equal to 2 sides, but most were bounded on less than 1 side.
- We eliminated some features that were likely only down-dropped blocks, and thus deemed more similar to chaos than a true pit as defined in the paper (example of such a feature at high res in Fig. 14).
- We applied a stricter limit on how well bounded a chaos feature must be. In some cases chaos is bounded by a scarp, trough, or obvious fracture. In other cases chaos grades into less fractured terrain. Sometimes it is ambiguous how to split or lump features, or one side is well bounded by a scarp or trough, but the other blends into the surrounding terrain (Greenberg et al., 1999; Collins and Nimmo, 2009).
- At the edges of the large areas mapped here as "regional, essentially interconnected chaos", there is a transition from a large interconnected mass of broken surface, to smaller individual upwellings, surrounded by unbroken ridged plains along the borders. This transitional contact is difficult to strictly define, but in the least conservative definition we counted more features as individual upwellings, while in this moderately conservative definition we counted more features as interconnected if there was any hint of continuous fracturing.
- We avoided defining a feature edge based on albedo alone. It is often the case that a "patch" of chaos will contain some dark terrain and also some terrain that is just as hummocky but not dark.

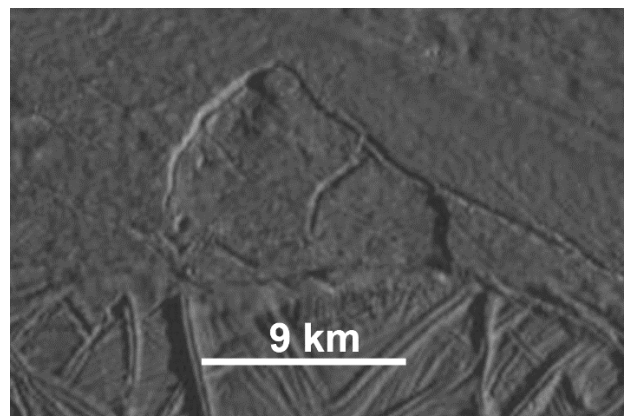

**Figure 14:** High resolution (68 m px$^{-1}$) example of down-dropped block.

*Most conservative definition:*
- Similar to moderately conservative definition, but we mapped only features bounded on at most 1 side by a ridge.
- Only very obvious features with well-defined boundaries were included. We did not, however, exclude any features solely based on small size.

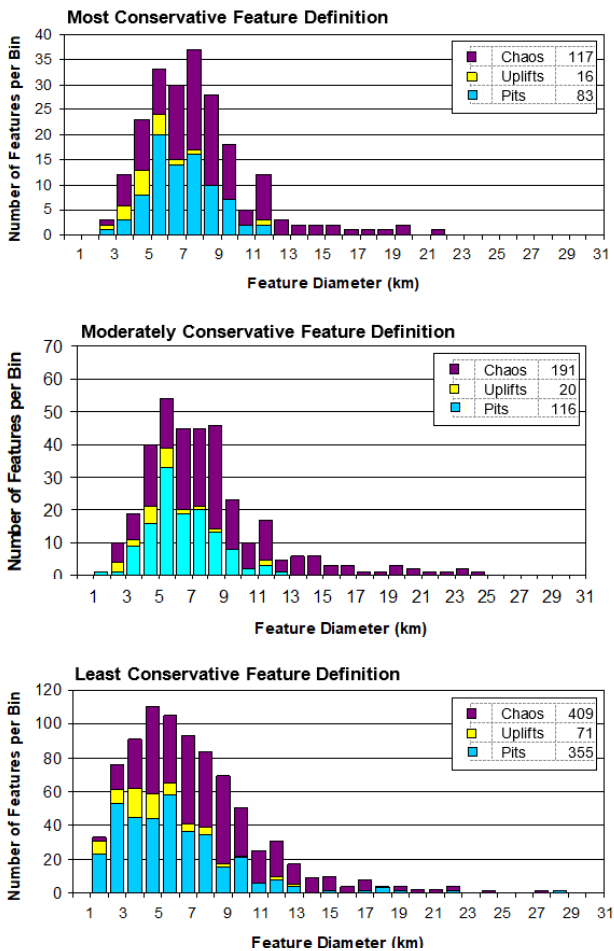

**Figure A.15: Results for more conservative versus more liberal mapping definitions.** Example are is the northern portion of the West RegMap (trailing hemisphere). Note changing scale on y-axis. The total number of features is shown for each mapping definition, as described in appendix text.

This experiment was conducted over the two most feature rich areas in the RegMaps: the southern part of the East RegMap and the northern portion of the West RegMap. The peak of the distribution remains relatively the same no matter what definition is used (Fig. A.15). The peak for the most conservative definition is the same as that for the moderate definition (between 5 and 7 km in diameter in a log-normal type distribution). The inclusion of everything that could be construed as a feature (least conservative definition) did shift the peak slightly; this distribution has a more distinct peak near 4-5 km in diameter. There is a significant drop off in the number of features below this peak, and especially below 2 km in diameter. This means we observed declining numbers of features above the practical resolution limit cutoff of ~1 km in diameter for a feature 5 pixels across. Because the two more conservative definitions are essentially subsets of least conservative definition, the depth-to-diameter and height-to-diameter trends are similar for all three (for the moderate definition these trends are shown in Fig. 9): increasing pit depths and uplift heights with increasing size, and the upper and lower bounds on the distributions remain unchanged.

## Appendix B: Comparison with Previous and Other Measurements

Collins and Nimmo (2009) compared several studies by plotting their results for feature diameter on a common scale (see their Fig. 12). The rebinned chaos data mapped by Riley et al. (2000) and Spaun (2002) match up fairly well in this plot (Collins and Nimmo, 2009; their Fig. 12a). There is more discrepancy between pits and uplifts mapped by Greenberg et al. (2003) within the RegMaps and domes mapped separately by Rathbun (1998) in the Conamara region. Rathbun (1998) found fewer small uplifts and more large ones, shifting their distribution and its peak towards larger sizes. We have rebinned and renormalized our data for pits to match the areal density bins of Greenberg et al. (2003) (Fig. B.16). We find substantially fewer small pits than Greenberg et al. (2003). Comparison of the maps provided in Greenberg et al. (2003) to those in this work show we mapped similar large features in many cases, but the hundreds of small pits implied in Greenberg et al. (2003) are difficult to discern at the resolution of their figures. In some cases these small features mapped by Greenberg et al. (2003) were considered in this study to be part of larger chaos areas, or were bounded on all sides by lineaments and thus considered to be only an apparent depressions created by the surrounding taller ridges. We attempted to limit ourselves to what could be verified with topography, and what could plausibly a single expression of an upwelling.

Comparing our data with those in the thesis work of Spaun (2002), as reported in Spaun et al. (2004), we find a favorable comparison (Fig. B.17). This plot is for all feature types mapped by this work on both RegMaps, and for all "micro-chaos lenticulae" and chaos features mapped by Spaun (2002). Spaun (2002) mapped some of the same areas as this study, but also other areas not included in this study. Given the differences in coverage and the fact that the two are completely independent mappings, there is a remarkable match between the peak in the distribution around features ~5 km in diameter. This reinforces the idea that all features may lie on a continuum of surface expressions from upwellings or disturbances in the shell, because the two studies focused on different features (although there is some overlap). The results presented here also compare favorably to the mean size for pits and uplifts reported in Culha and Manga (2016), although they mapped fewer pits than this study. The pits and uplifts mapped here are those that were verified with topographic data. Noviello et al. (2019) found similar size distributions to that presented here, and also found that pits were the most common sub-circular feature that was not chaotic/broken terrain.

A spectacular high-resolution mosaic was obtained by *Galileo* within the northern West

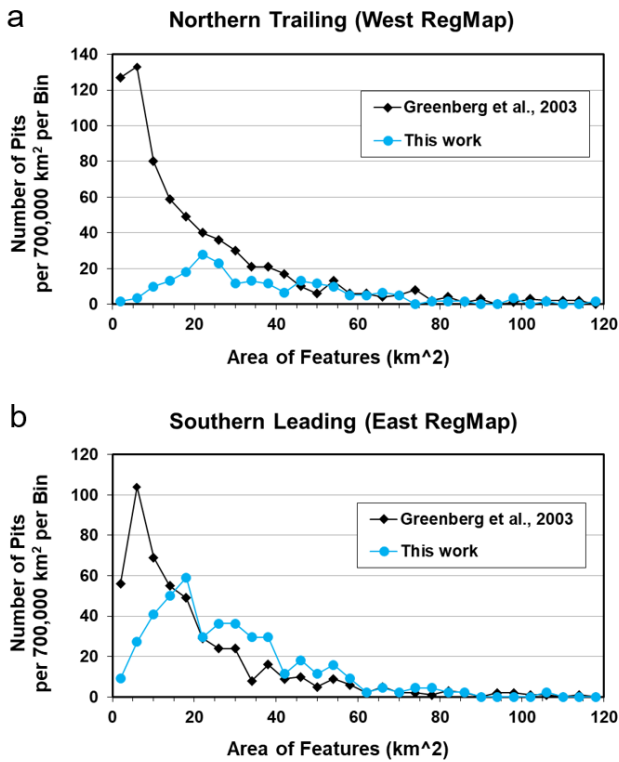

**Fig. B.16: Comparison with Greenberg et al. 2003.** (a) West RegMap (northern section only) feature areas (bins as in Greenberg et al., 2003). (b) East RegMap (southern section only) feature areas. We normalized data from this work by the area of the northern trailing (~4.3 x $10^5$ km$^2$) or southern leading (~3.1 x $10^5$ km$^2$) RegMaps where features were mapped (i.e., this area does not include the large areas of regional chaos). A severe mismatch in the number of small features is evident for both RegMaps.

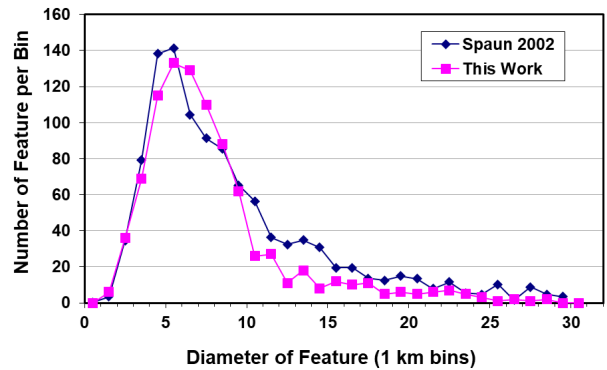

**Figure B.17: Comparison with Spaun 2002.** Data from this work and Spaun (2002) both show a peak in the size-frequency distribution of features near 5-6 km in diameter.

RegMap (see Fig. 1) during Europa encounter E19 (19ESRHADAM01). Centered near 34°N, 223°W, just north of Rhadamanthys Linea, it was viewed by Galileo at low sun during E19 at 63 m/pixel and E15 at 220 m/pixel, allowing the creation of stereo-controlled PC DEM with 63-m horizontal resolution, 45-m absolute (and several meter relative) vertical resolution and reliable regional slope control (Fig. B1.8). Examined in part in the main text, an overview of the topography in this region reveals eight prominently deep and well-defined pits, as well as several less pronounced depressions (Fig. B.18b) and subtler structures. Although some of the pits are clustered, others occur singly and are more widely spaced at distances of up to ~50 km. The pits at the Rhadamanthys site range in length from 5 to 15 km. From our stereo-controlled PC DEM, we find that the depths of the 8 largest pits are between 100 and 450 m deep. Several chaos features are also observed within the site, which show complete or near-complete resurfacing of the pre-existing terrain. Although similar in size to neighboring pits, they differ in that they are clearly formed of a darker rugged or blocky material (Greenberg et al., 1999; Collins and Nimmo, 2009). In our DEM these features either have no relief or are elevated only up to 50 m relative to local mean height. *Most importantly, it is evident that there is no population of much smaller pits, uplift, and chaos.*

Thus, one can be forgiven, when examining the Rhadamanthys site, in concluding that there might be a relatively large characteristic or modal size for pits, uplifts, and small-scale chaos. Early *Galileo* studies, also based on limited regions (Pappalardo et al., 1998; Rathbun et al., 1998) came to such conclusions. Our work shows that on a more global scale that ~5-6 km equivalent diameter is a better modal size for these features than ~10 km. But as the physical drivers for the formation of these features may themselves have a local or regional character (in space and time), it is probably important to consider feature characteristics on a region-to-region basis when modeling and developing interpretations.

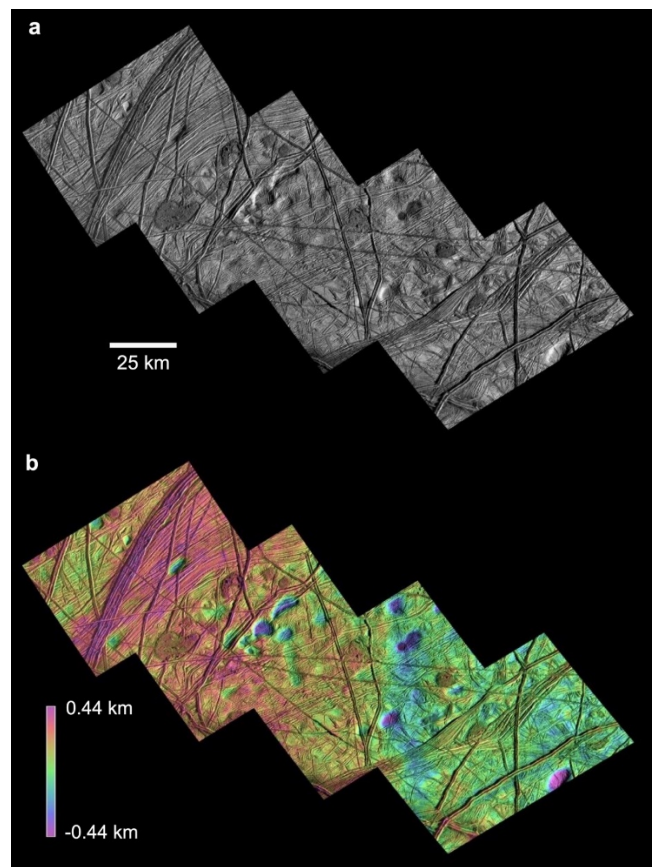

Fig. B.18. **High resolution view of the Rhadamanthys Region.** (a) Base mosaic (65 m px$^{-1}$), and (b) stereo-controlled photoclinometric digital elevation model. A fundamental characteristic of the pits here (and in the REGMAP areas as well) is that the ridged plains in which they occur are clearly undisturbed and continuous across depressions. Thus, pits as such have not been resurfaced, nor has the fine-scale ridge topography within them been relaxed or removed. Interior wall slopes can vary, depending on pit depth, but for deeper examples, wall slopes are typically 7-10° and can be as high as 20°. Pits also conspicuously lack raised rims. In some cases, pit floors slope asymmetrically to one side, but the majority appear flat-floored, except for the ubiquitous finer-scale ridges. See also close-ups in Figs. 5a and 7b.

# Appendix C: Density Estimates for Europa's Ice Shell and Ocean

Equation (1) is plotted in Fig. 13 (solid black line) for the simple case of solid water ice floating in pure liquid water ($\rho_l$ = 1000 kg m$^{-3}$ and $\rho_s$ = 929 kg m$^{-3}$; Table C.1). The ice density has been adjusted for an average temperature $T$ of 170 K, assuming a conductive temperature profile based on temperature dependent ice conductivity $k = 567/T$ W m$^{-1}$ (Klinger, 1980) and density (Petrenko and Whitworth, 1999). As can be seen, maximum pit depths of 400 to 500 m, as are observed on Europa, translate into minimum shell thicknesses of ~6 km for pure ice over water. Conductive profiles might be preferred over convective ones as we are determining minimum shell thicknesses (least favorable for convection). Nevertheless, convection cannot be ruled out, either during or after pit formation, so we illustrate this case as well in Fig. 13 (dashed black line). For a largely convecting ice shell, $T$ ~260 K (McKinnon, 1999) and $\rho_s \approx$ 919 kg m$^{-3}$, so the minimum shell thicknesses are 12% less).

Neither the shell nor the ocean beneath it are pure H$_2$O in composition (Zolotov and Kargel, 2009). A substantial non-ice component has been detected in *Galileo* NIMS and Earth-based spectra of stratigraphically young and visually darkened, yellow-brown regions. On this topic, see review in Carlson et al. (2009) and numerous subsequent papers (e.g., Hand and Carlson, 2015; Trumbo et al., 2019). As for the ocean, the most direct observational constraints are from the induced magnetic field measurements from the *Galileo* magnetometer. As summarized in McKinnon et al. (2009), the most complete simulation to date of Europa's magnetospheric interaction is that of Schilling et al. (2007). They find that the product of ocean conductivity and thickness must exceed 50 S m$^{-1}$ km. For an ocean thickness of 100 km and a 25-km-thick ice shell, the lower limit on ocean conductivity is 0.5 S m$^{-1}$, and for thinner ice shells the conductivity limit is lower still. An upper limit on the oceanic conductivity cannot be set with the available data, but the lower limit is well below that of the cold terrestrial sea water value of 3.2 S m$^{-1}$ (Tyler et al., 2017) or the saturated conductivity of a H$_2$O-MgSO$_4$ brine at 6 S m$^{-1}$ (Hand and Chyba, 2007). That is, the magnetometer measurements imply that Europa's ocean contains an electrolyte, but the ocean is not required to be as salty as the Earth's ocean. But it may be.

With these compositional constraints in mind, we broaden the range of density differences ($\Delta\rho$) under consideration. Because we only have lower limits on ocean conductivity, we look at extremes of possible density variation, which are provided by sulfate salts[5]. We take the dominant non-ice (salt) components of the ocean to be MgSO$_4$ and Na$_2$SO$_4$. If the sulfate concentration is in the water-ice-rich corner with respect to the H$_2$O-MgSO$_4$-Na$_2$SO$_4$ ternary eutectic at the lowest melting point (Kargel, 1991), then pure ice will freeze from the top down, expelling denser brine as it does, as occurs on Earth (e.g., Buffo et al., 2020). The ternary eutectic liquid has a density of 1208 kg m$^{-3}$ (Kargel et al., 2000, their Table II), so the maximum density difference between the ocean and solid ice is relatively large, which translates into a lower bound to the minimum shell thicknesses (Fig. 13). In contrast, hypersaline sulfate oceans first precipitate sulfate layers on the ocean floor (Kargel, 1991; Kargel et al., 2000; Spaun and Head, 2001) and the ocean composition is driven toward the ternary eutectic as the temperature drops. At the eutectic temperature (268 K), a buoyant surface shell of eutectic composition can begin to freeze out, composed of 51.3 wt% ice, 42.3% MgSO$_4$·11H$_2$O (meridianiite), and 6.4% Na$_2$SO$_4$·10H$_2$O (mirabilite). With a density of 1134 kg m$^{-3}$ (cf. Kargel et al., 2000) it is buoyant with respect to eutectic liquid, because it is half water ice by mass, and the smaller $\Delta\rho$ implied yields an upper bound to the minimum shell thicknesses, for a given pit depth.

---

[5] It is difficult cosmochemically to achieve high oceanic salt concentrations with magnesium or alkali halides alone because of aqueous dilution (Zolotov and Shock, 2001). That is, chorine, bromine, etc., are only so cosmochemically abundant. This is not an issue for sulfur or sulfates.

**Table C.1: Ice Shell and Ocean Densities**

| Material Composition and Phase | Temperature (K) | Density (kg m$^{-3}$) |
|---|---|---|
| Water (liquid) | 273 | 1000 |
| Conductive water ice | 170 | 929 |
| Convecting water ice | 260 | 919 |
| H$_2$O-Meridianiite-Mirabilite eutectic liquid | 268 | 1208 |
| H$_2$O-Meridianiite-Mirabilite eutectic solid | 180 | 1150 |
| H$_2$O-Epsomite-Mirabilite eutectic liquid | 266 | 1220 |
| H$_2$O-Epsomite-Mirabilite eutectic solid | 180 | 1140 |

Note that we have corrected the solid eutectic composition in Kargel et al. (2000) for meridianiite (as opposed to the dodecahydrate) and a cooler, conductive shell temperature, using the unit cell values in Fortes et al. (2008) for meridianiite and those in Brand et al. (2009) for mirabilite, giving a $\rho_s \approx 1150$ kg m$^{-3}$ (Table C.1) in Fig. 13. And if for some reason meridianiite crystallization is kinetically inhibited (Kargel, 1991), the eutectic liquid composition should shift to be slightly more sulfate rich (near 18 wt% MgSO$_4$ and 2% Na$_2$SO$_4$), and the coexisting solid phase would be ~58% ice, 37% epsomite, and 5% mirabilite. The liquid and solid densities at this eutectic are ≈1220 and 1130 kg m$^{-3}$, which (with temperature correction) leaves the lower limit curve practically unchanged while reducing the upper limit curve (Fig. 13, thin line).

*Clathrates?* Hand et al. (2006) suggested that Europa's bulk ice shell may be substantially clathrated, with several percent by number O$_2$ (mainly), CO$_2$, and SO$_2$ produced radiolytically at the surface ultimately ending up distributed throughout the shell. Although the mechanisms for this distribution are uncertain, the important point here is that all these guest molecules increase act to increase the density of water ice irrespective of the composition of the underlying ocean. Thus, $\Delta\rho$ in Eq. (1) would be lessened, and $H$ values increased, which emphasizes that the limiting values in Fig. 13 are minimum values.

*Crustal Porosity.* Finally, we have not attempted to account for any porosity in the shell immediately surrounding the pits. Given the essentially igneous origin for Europa's shell, likely local "high-temperature" metamorphism during pit formation, and the modest role of cratering on Europa in creating a regolith, we feel neglecting porosity is justified to first order (and it is poorly constrained in any case). Even on the Moon, where the ancient crust has been subject to a ferocious impact bombardment, the porosity of the bulk crust is limited to 12% (Wieczorek et al., 2013). Accounting for (e.g.) 300-500 m deep pits on Europa by porosity loss alone (as proposed for depressions on Enceladus by Schenk and McKinnon (2009)) would require differential loss of such porosity over a vertical ice column of 2.5-4.2 km. Lower, and arguably more likely, initial porosities in Europa's elastic lithosphere would require even greater vertical extents of differential compaction.

## Appendix D: Supplementary data files

Both comma-delimited (.csv, 56 KB) and Excel (.xlsx, 83 KB) files with the feature measurements are available in an open online archive at https://doi.org/10.6084/m9.figshare.13515377. These files contain the following information for each feature: feature id, type (pit, uplift, or chaos), effective diameter (described above), location (longitude and latitude in degrees in a west positive longitude system), the feature depth/height for pits/uplifts. Additionally, there are notes for a few a few features, including noting transitional features (e.g., if a feature was similar to both the uplift and chaos categories). The region for each feature is also noted, where SL = southern leading, NL = northern leading, ST = southern trailing, and NT = northern trailing. Note that the feature ids are unique for each region but may repeat between regions.

# References

Barr, A.C., Showman, A.P., 2009. Heat transfer in Europa's icy shell. In: Pappalardo, R. T., McKinnon, W. B., Khurana, K. K., (Eds.), Europa. University of Arizona Press, Tucson, pp. 405-430.

Bierhaus, E.B., Zahnle, K., Chapman, C.R., 2009. Europa's crater distributions and surface ages. In: Pappalardo, R. T., McKinnon, W. B., Khurana, K. K., (Eds.), Europa. Univ. Arizona Press, Tucson, pp. 161-180.

Brand, H.E.A., Fortes, A.G., Wood, I.G., Knight, K.S., Volčaldo, L., 2009. The thermal expansion and crystal structure of mirabilite ($Na_2SO_4 \cdot 10D_2O$) from 4.2 to 300 K, determined by time-of-flight neutron powder diffraction. Phys. Chem. Minerals 36, 29–46.

Buffo, J.J., Schmidt, B.E., Huber, C., Walker, C.C., 2020. Entrainment and dynamics of ocean-derived impurities within Europa's ice shell. J. Geophys. Res. Planets 125, e06394. doi:10.1029/2020je006394

Carlson, R.W., et al., 2009. Europa's surface composition. In: Pappalardo, R. T., McKinnon, W. B., Khurana, K. K., (Eds.), Europa. University of Arizona Press, Tucson, pp. 283-328.

Collins, G.C., Head, J.W., Pappalardo, R.T., Spaun, N.A., 2000. Evaluation of models for the formation of chaotic terrain on Europa. J. Geophys. Res. 105, 1709-1716.

Collins, G.C., Nimmo, F., 2009. Chaotic terrain on Europa. In: Pappalardo, R. T., McKinnon, W. B., Khurana, K. K., (Eds.), Europa. Univ. Arizona Press, Tucson, pp. 259-282.

Cox, R., Bauer, A.W., 2015. Impact breaching of Europa's ice: Constraints from numerical modeling. J. Geophys. Res. Planets 120, 1708. doi:10.1002/2015je004877

Craft, K.L., Patterson, G.W., Lowell, R.P., Germanovich, L., 2016. Fracturing and flow: Investigations on the formation of shallow water sills on Europa. Icarus 274, 297-313. doi:10.1016/j.icarus.2016.01.023

Craft, K.L., Walker, C.C., Quick, L.C., Lowell, R.P., 2019. "Freckles," "spots," and domes on Europa and Ceres: Surface features driven by subsurface cryovolcanic diking, and surface response? Lunar and Planetary Science Conference 50. 3102 (abstract).

Culha, C., Manga, M., 2016. Geometry and spatial distribution of lenticulae on Europa. Icarus 271, 49-56.

Doggett, T., Greeley, R., Figueredo, P., Tanaka, K., 2009. Geologic stratigraphy and evolution of Europa's surface. In: Pappalardo, R. T., McKinnon, W. B., Khurana, K. K., (Eds.), Europa. University of Arizona Press, Tucson, pp. 137-160.

Dombard, A.J., McKinnon, W.B., 2006. Folding of Europa's icy lithosphere: An analysis of viscous-plastic buckling and subsequent topographic relaxation. J. Struct. Geol. 28, 2259-2269.

Fagents, S.A., 2003. Considerations for effusive cryovolcanism on Europa: The post-Galileo perspective. J. Geophys. Res. Planets 108, 5139.

Figueredo, P.H., Greeley, R., 2004. Resurfacing history of Europa from pole-to-pole geologic mapping. Icarus 167, 287-312.

Fortes, A.D., Wood, I.G., Knight, K.S., 2008. The crystal structure and thermal expansion tensor of $MgSO_4 \cdot 11D_2O$ (meridianiite) determined by neutron powder diffraction. Physics and Chemistry of Minerals 35, 207. doi:10.1007/s00269-008-0214-x

Galland, O., Planke, S., Neumann, E.-R., Malthe-Sørenssen, A., 2009. Experimental modelling of shallow magma emplacement: Application to saucer-shaped intrusions. Earth Planet. Sci. Lett. 277, 373. doi:10.1016/j.epsl.2008.11.003

Goodman, J.C., Collins, G.C., Marshall, J., Pierrehumbert, R.T., 2004. Hydrothermal plume dynamics on Europa: Implications for chaos formation. J. Geophys. Res. 109, 03008.

Greeley, R., et al., 2000. Geologic mapping of Europa. J. Geophys. Res. 105, 22559-22578.

Greeley, R., Pappalardo, R.T., Prockter, L.M., Hendrix, A.R., Lock, R.E., 2009. Future exploration of Europa. In: Pappalardo, R. T., McKinnon, W. B., Khurana, K. K., (Eds.), Europa. Univ. Arizona Press, Tucson, pp. 655-696.

Greenberg, R., Geissler, P., 2002. Europa's dynamic icy crust. Meteorit. Planet. Sci. 37, 1685. doi:10.1111/j.1945-5100.2002.tb01158.x

Greenberg, R., Hoppa, G.V., Tufts, B.R., Geissler, P., Riley, J., Kadel, S., 1999. Chaos on Europa. Icarus 141, 263-286.

Greenberg, R., Leake, M.A., Hoppa, G.V., Tufts, B.R., 2003. Pits and uplifts on Europa. Icarus 161, 102-126.

Han, L., Showman, A.P., 2005. Thermo-compositional convection in Europa's icy shell with salinity. Geophys. Res. Lett. 32, 20201.

Hand, K.P., Carlson, R.W., 2015. Europa's surface color suggests an ocean rich with sodium chloride. Geophys. Res. Lett. 42, 3174. doi:10.1002/2015gl063559

Hand, K.P., Chyba, C.F., 2007. Empirical constraints on the salinity of the europan ocean and implications for a thin ice shell. Icarus 189, 424. doi:10.1016/j.icarus.2007.02.002

Hand, K.P., Chyba, C.F., Carlson, R.W., Cooper, J.F., 2006. Clathrate hydrates of oxidants in the ice shell of Europa. Astrobiology 6, 463. doi:10.1089/ast.2006.6.463

Hand, K.P., Chyba, C.F., Priscu, J.C., Carlson, R.W., Nealson, K.H., 2009. Astrobiology and the potential for life on Europa. In: Pappalardo, R. T., McKinnon, W. B., Khurana, K. K., (Eds.), Europa. Univ. Arizona Press, Tucson, pp. 589-630.

Head, J.W., Sherman, N.D., Pappalardo, R.T., Thomas, C., Greeley, R., Team, G.S., 1998. Cryovolcanism on Europa: Evidence for the emplacement of flows and related deposits in the E4 region (5N, 305W) and interpreted eruption conditions. Lunar Planet. Sci. XXIX, 1491 (abstract).

Hemingway, D., Iess, L., Tajeddine, R., Tobie, G., 2018. The interior of Enceladus. In: Schenk, P. M., Clark, R. N., Howett, C. J. A., Verbiscer, A. J., Waite, J. H., (Eds.), Enceladus and the Icy Moons of Saturn. University of Arizona Press, Tucson, pp. 57-77. doi:10.2458/azu_uapress_9780816537075-ch004

Hoppa, G.V., Greenberg, R., Riley, J., Tufts, B.R., 2001. Observational selection effects in Europa image data: Identification of chaotic terrain. Icarus 151, 181-189.

Howell, S.M., Smith, M., Otis, R., 2020. Europa's likely icy thickness: Pulling ice shells out of a hat. Lunar Planet. Sci. 51, 2957 (abstract).

Kargel, J.S., 1991. Brine volcanism and the interior structures of asteroids and icy satellites. Icarus 94, 368. doi:10.1016/0019-1035(91)90235-l

Kargel, J.S., et al., 2000. Europa's crust and ocean: Origin, composition, and the prospects for life. Icarus 148, 226. doi:10.1006/icar.2000.6471

Kattenhorn, S.A., Hurford, T., 2009. Tectonics of Europa. In: Pappalardo, R. T. M., W.B., Khurana, K. K., (Eds.), Europa. University of Arizona Press, Tuscon, pp. 199-236.

Kattenhorn, S.A., Prockter, L.M., 2014. Evidence for subduction in the ice shell of Europa. Nature Geoscience 7, 762–767. doi:10.1038/ngeo2245

Klinger, J., 1980. Influence of a phase transition of ice on the heat and mass balance of comets. Science 209, 271. doi:10.1126/science.209.4453.271

Lainey, V., Arlot, J.-E., Karatekin, Ö., van Hoolst, T., 2009. Strong tidal dissipation in Io and Jupiter from astrometric observations. Nature 459, 957-959.

Leonard, E.J., Patthoff, D.A., Senske, D.A., Collins, G.C., Bunte, M.K., Doggett, T., 2017. Updating the global geologic map of Europa. Third Planetary Data Workshop and The Planetary Geologic Mappers Annual Meeting. 1986 (abstract).

Lesage, E., Massol, H., Schmidt, F., 2020. Cryomagma ascent on Europa. Icarus 335, 113369. doi:10.1016/j.icarus.2019.07.003

Lucchitta, B.K., Soderblom, L.A., 1982. The geology of Europa. In: Morrison, D., (Ed.), The satellites of Jupiter University of Arizona Press, Tucson, pp. 521-555.

Manga, M., Michaut, C., 2017. Formation of lenticulae on Europa by saucer-shaped sills. Icarus 286, 261-269. doi:10.1016/j.icarus.2016.10.009

McKinnon, W.B., 1999. Convective instability in Europa's floating ice shell. Geophys. Res. Lett. 26, 951-954.

McKinnon, W.B., Pappalardo, R.T., Khurana, K.K., 2009. Europa: Perspectives on an ocean world. In: Pappalardo, R. T., McKinnon, W. B., Khurana, K. K., (Eds.), Europa. Univ. Arizona Press, Tucson, pp. 697-710.

McKinnon, W.B., Zolensky, M.E., 2003. Sulfate content of Europa's ocean and shell: Evolutionary considerations and some geological and astrobiological implications. Astrobiology 3, 879-897.

Michaut, C., Manga, M., 2014. Domes, pits, and small chaos on Europa produced by water sills. J. Geophys. Res. Planets 119, 550-573.

Mitri, G., Showman, A.P., 2008. A model for the temperature-dependence of tidal dissipation in convective plumes on icy satellites: Implications for Europa and Enceladus. Icarus 195, 758. doi:10.1016/j.icarus.2008.01.010

Neish, C., Prockter, L.M., Patterson, G.W., 2011. The identification of chaotic terrain on Europa. EPSC-DPS2011. 6, 259 (abstract).

Nimmo, F., Giese, B., Figueredo, P., Moore, W.B., 2004. Thermal and topographic tests of Europa chaos formation models. Lunar Planet. Sci. XXXV, 1403 (abstract).

Nimmo, F., Manga, M., 2009. Geodynamics of europa's icy shell. In: Pappalardo, R. T., McKinnon, W. B., Khurana, K. K., (Eds.), Europa. Univ. Arizona Press, Tucson, pp. 381-404.

Noviello, J.L., Torrano, Z.A., Rhoden, A.R., Singer, K.N., 2019. Mapping Europa's microfeatures in regional mosaics: New constraints on formation

models. Icarus 329, 101. doi:10.1016/j.icarus.2019.02.038

O'Brien, D.P., Geissler, P., Greenberg, R., 2002. A melt-through model for chaos formation on Europa. Icarus 156, 152-161.

Pappalardo, R.T., Barr, A.C., 2004. The origin of domes on Europa: The role of thermally induced compositional diapirism. Geophys. Res. Lett. 31, 01701.

Pappalardo, R.T., et al., 1998. Geological evidence for solid-state convection in Europa's ice shell. Nature 391, 365-368.

Peddinti, D.A., McNamara, A.K., 2019. Dynamical investigation of a thickening ice-shell: Implications for the icy moon Europa. Icarus 329, 251–269. doi:10.1016/j.icarus.2019.03.037

Petrenko, V.F., Whitworth, R.W., 1999. Physics of ice. Oxford Univ. Press, Oxford, UK.

Polteau, S., Mazzini, A., Galland, O., Planke, S., Malthe-Sørenssen, A., 2008. Saucer-shaped intrusions: Occurrences, emplacement and implications. Earth Planet. Sci. Lett. 266, 195. doi:10.1016/j.epsl.2007.11.015

Quick, L.C., Glaze, L.S., Baloga, S.M., 2017. Cryovolcanic emplacement of domes on Europa. Icarus 284, 477-488. doi:10.1016/j.icarus.2016.06.029

Rathbun, J.A., Musser, G.S., Squyres, S.W., 1998. Ice diapirs on Europa: Implications for liquid water. Geophys. Res. Lett. 25, 4157-4160.

Riley, J., Hoppa, G.V., Greenberg, R., Tufts, B.R., Geissler, P., 2000. Distribution of chaotic terrain on Europa. J. Geophys. Res. 105, 22599-22616.

Schenk, P.M., 2002. Thickness constraints on the icy shells of the Galilean satellites from a comparison of crater shapes. Nature 417, 419-421.

Schenk, P.M., McKinnon, W.B., 2001. Topographic variability on Europa from Galileo stereo images. Lunar Planet. Sci. XXXII, 2078 (abstract).

Schenk, P.M., McKinnon, W.B., 2009. One-hundred-km-scale basins on Enceladus: Evidence for an active ice shell. Geophys. Res. Lett. 36, L16202. doi:10.1029/2009gl039916

Schenk, P.M., Pappalardo, R.T., 2004. Topographic variations in chaos on Europa: Implications for diapiric formation. Geophys. Res. Lett. 31, 16703.

Schenk, P.M., Wilson, R.R., Davies, A.G., 2004. Shield volcano topography and the rheology of lava flows on Io. Icarus 169, 98-110.

Schilling, N., Neubauer, F.M., Saur, J., 2007. Time-varying interaction of Europa with the jovian magnetosphere: Constraints on the conductivity of Europa's subsurface ocean. Icarus 192, 41. doi:10.1016/j.icarus.2007.06.024

Schmidt, B.E., Blankenship, D.D., Patterson, G.W., Schenk, P.M., 2011. Active formation of 'chaos terrain' over shallow subsurface water on Europa. Nature 479, 502-505.

Senske, D.A., Leonard, E.J., Patthoff, D.A., Collins, G.C., 2018. The Europa global geologic map. Lunar Planet. Sci. 49, 1340 (abstract).

Showman, A.P., Han, L., 2004. Numerical simulations of convection in Europa's ice shell: Implications for surface features. J. Geophys. Res. Planets 109, 01010. doi:10.1029/2003JE002103

Showman, A.P., Han, L., 2005. Effects of plasticity on convection in an ice shell: Implications for Europa. Icarus 177, 425-437.

Singer, K.N., 2013. Icy satellite tectonic, geodynamic, and mass wasting surface features: Constraints on interior processes and evolution. Thesis, Washington University in St. Louis, St. Louis. 220 pp.

Singer, K.N., Bland, M.T., Schenk, P.M., McKinnon, W.B., 2018. Relaxed impact craters on Ganymede: Regional variation and high heat flows. Icarus 306, 214-224. doi:10.1016/j.icarus.2018.01.012

Singer, K.N., McKinnon, W.B., Schenk, P.M., 2010. Pits, spots, uplifts, and small chaos regions on Europa: Evidence for diapiric upwelling from morphology and morphometry. Lunar Planet. Sci. 41, 2195 (abstract).

Singer, K.N., McKinnon, W.B., Schenk, P.M., Moore, J.M., 2012. Massive ice avalanches on Iapetus mobilized by friction reduction during flash heating. Nature Geoscience 5, 574-578.

Skjetne, H.L., et al., 2021. Morphological comparison of blocks in chaos terrains on Pluto, Europa, and Mars. Icarus 356, 113866. doi:10.1016/j.icarus.2020.113866

Sotin, C., Head, J.W., Tobie, G., 2002. Europa: Tidal heating of upwelling thermal plumes and the origin of lenticulae and chaos melting. Geophys. Res. Lett. 29, 74-1.

Sotin, C., Tobie, G., Wahr, J., McKinnon, W.B., 2009. Tides and tidal heating on Europa. In: Pappalardo, R. T., McKinnon, W. B., Khurana, K. K., (Eds.), Europa. Univ. Arizona Press, Tucson, pp. 85-117.

Spaun, N.A., 2002. Chaos, lenticulae, and lineae on Europa: Implications for geological history, crustal thickness, and the presence of an ocean. Thesis, Brown University, Providence. 334 pp.

Spaun, N.A., Head, J.W., 2001. A model of Europa's crustal structure: Recent Galileo results and

implications for an ocean. J. Geophys. Res. 106, 7567. doi:10.1029/2000je001270

Spaun, N.A., Head, J.W., Pappalardo, R.T., 2004. Europan chaos and lenticulae: A synthesis of size, spacing, and areal density analyses. Lunar Planet. Sci. XXXV, 1409 (abstract).

Spaun, N.A., Pappalardo, R.T., Head, J.W., 2001. Equatorial distribution of chaos and lenticulae on Europa. Lunar Planet. Sci. XXXII, 2132 (abstract).

Steinbrügge, G., et al., 2020. Brine migration and impact-induced cryovolcanism on Europa. Geophys. Res. Lett. 47, e90797. doi:10.1029/2020gl090797

Trompert, R.A., Hansen, U., 1998. On the Rayleigh number dependence of convection with a strongly temperature-dependent viscosity. Physics of Fluids 10, 351. doi:10.1063/1.869527

Trumbo, S.K., Brown, M.E., Hand, K.P., 2019. Sodium chloride on the surface of Europa. Science Advances 5, aaw7123. doi:10.1126/sciadv.aaw7123

Tyler, R.H., Boyer, T.P., Minami, T., Zweng, M.M., Reagan, J.R., 2017. Electrical conductivity of the global ocean. Earth, Planets, and Space 69, 156. doi:10.1186/s40623-017-0739-7

Wieczorek, M.A., et al., 2013. The crust of the Moon as seen by GRAIL. Science 339, 671-675. doi:10.1126/science.1231530

Williams, K.K., Greeley, R., 1998. Estimates of ice thickness in the Conamara Chaos region of Europa. Geophys. Res. Lett. 25, 4273-4276.

Zolotov, M.Y., Kargel, J.S., 2009. On the chemical composition of Europa's icy shell, ocean, and underlying rocks. In: Pappalardo, R. T., McKinnon, W. B., Khurana, K. K., (Eds.), Europa. University of Arizona Press, Tucson, pp. 431-458.

Zolotov, M.Y., Shock, E.L., 2001. Composition and stability of salts on the surface of Europa and their oceanic origin. J. Geophys. Res. 106, 32815. doi:10.1029/2000je001413